\magnification=\magstep1
\overfullrule=0pt

\vglue 1truein

\centerline{\bf Marietta Blau's Work After World War II}
\centerline{\bf Arnold Perlmutter}
\centerline{\bf Department of Physics}
\centerline{\bf University of Miami}
\centerline{\bf Coral Gables, Florida 33124}
\vskip 10pt
\centerline{\bf completed, October 27, 2000}
\vskip 24pt
\noindent
This paper has been translated into German and will be included, in a
somewhat altered form, in a book {\it Sterne der Zertrummerung, Marietta
Blau, Wegbereiterin der Moderne Teilchenphysik}, Brigitte Strohmaier and
Robert Rosner, eds., Boehlau Verlag, Wien.

\vfill\eject 
\vglue 1truein
\vskip 16pt
\leftline{A. Introduction}
\vskip 12pt
While it is clear that the seminal work of Dr. Marietta Blau was done
in the 1920's and especially in the 1930's, it is also evident that her
separation from the great research centers from 1938 to 1944 had a
devastating effect on her productivity.  It was during this period that
Cecil F. Powell, at Bristol University, made use of Blau's earlier
tutelage on the preparation and analysis of photographic emulsions.
According to Blau's conversations with me (much later), she consulted
with Ilford in the 1930's to improve emulsion sensitivity and
uniformity, and presumably had also imparted crucial lore of the technique
to Powell.  C.F. Powell, who had been a student of C.T.R. Wilson, had
employed cloud chambers in a wide variety of studies in vulcanology,
mechanical engineering, and nuclear physics.  In 1938 and 1939 Powell's
experimental efforts turned to the use of photographic emulsions to
investigate neutron interactions, and then to nuclear reactions$^{1}$.
With the coming of the war and then the British nuclear atomic bomb
project, Powell established a formidable laboratory and collaboration
for the analysis of emulsions and for their improvement by Ilford and
Kodak.  Thus it came about that Powell and his collaborators discovered
the pion in emulsions, exposed in 1947 at high altitudes in the Bolivian
Andes and Pyrennes.  Powell then received the Nobel Prize for Physics in
1950``for his development of the photographic method of studying nuclear
processes and for the resulting discovery of the pion (pi-meson), a
heavy subatomic particle.''
\vskip 12pt
It stretches one's credibility that Blau should not have shared in the
first part of the citation, if not for the prejudices or
narrowmindedness of the Swedish Academy, demonstrated on several other
occasions (at least in the cases of Lise Meitner and C.S. Wu).  The great
Erwin Schr\"odinger himself twice nominated Blau for a Nobel Prize
$^2$.
There certainly may have been other nominations.  Consider, for example,
this quote from the classic text on atomic physics by Max Born
$^3$:
\vskip 12pt
``Another great advance was made by two Viennese ladies, Misses Blau and
Wambacher (1937), who discovered a photographic method of recording
tracks of particles.  The grains of emulsion are sensitive not only to
light but also to fast particles; if a plate exposed to a beam of
particles is developed and fixed the tracks are seen under the
microscope as chains of black spots.  Their quality depends very much on
the size of the grains, and special emulsions with very small and dense
grains have been developed (Ilford, Kodak).
\vskip 12pt
``The photographic tracks are some thousand times shorter than
corresponding tracks in air, because of the higher stopping power of the
solid material; they are of the order of some microns.  The advantages
of this method are its extreme simplicity, the continuity of
sensitiveness, and the great number of events recorded on one plate.  On
the other hand, high-quality tracks are observed and micro-photographed
with oil-immersion objectives which have a narrow depth of focus; hence
only a restricted part of a track appears sharp, and several photographs
have to be taken with different focus.''
\vskip 12pt
My own knowledge of Marietta Blau's post war research is based somewhat
on my personal contacts with her during 1956-1963 in Miami and Vienna
(as well as with some
of her earlier colleagues), but mainly from reading her papers written
after 1945, and from several informative references.  The incisive book
by Peter Galison, {\it Image and Logic, A Material Culture of
Microphysics}$^1$, and
a subsequent article in {\it Physics Today}$^4$ contain much useful information on Blau's
life.  Other sources include Leopold Halpern's biographical sketch
$^5$, the internet web-page by Nina Byers,
$^6$ and the elaborate volume
prepared by C.F. Powell, P.H. Fowler and D.H. Perkins. $^7$
\vskip 12pt
As I stated above, Blau's six year residence in Mexico effectively
removed her from serious research.  Galison states that she had to teach
24 hours a week, and in addition suffered the
theft of materials that would have allowed her to establish a research
laboratory.$^1$  I do not know much about her years in Mexico, except for
recalling that she worked hard to support her infirm mother, and that in
the circle of European intellectuals that she frequented was also the
famous exile, Leon Trotsky.  In that group was also an erratic young man
who was later identified as Trotsky's assassin.  Marietta said that she
and her friends attempted to warn Trotsky of the man's dangerousness,
but that he dismissed their entreaties, and was murdered in 1940.  The
prevalent view is that the assassin was a Stalin agent, but I have no
other information as to its veracity.
\vskip 12pt
\leftline{B.  The First Scintillation Counter}
\vskip 12pt
I do not know of the circumstances that brought her to New York in 1944,
when she first went to work for the International Rare Metals Refinery
and later the Canadian Radium and Uranium Corporation.  What is clear is
that her frustration at being pent up in scientifically remote Mexico
led to an explosion of creative activity, in spite of the fact that she
at first found herself at the periphery of the American research
establishment.
\vskip 12pt
Blau's first paper published after she came to the United States was in
1945, with B. Dreyfus $^{B46}$.  As far as I can tell, it was
the first example to be published in the open literature on the use of
the photomultiplier tube in conjunction with scintillating target, a $ZnS$
screen to detect radioactive emissions.  They actually measured the
phototube current as a function of distance between the
$\alpha$-particle source and the phototube, and observed a clear inverse
square-law dependence.  The paper is striking in its straightforwardness
and simplicity.
\vskip 12pt
According to the book by J.B. Birks$^8$, the device was actually used as a dosimeter but it can
be regarded as the first rudimentary scintillation counter, a great
advance over manual counting of light flashes by human observers, as
pioneered by Rutherford and his collaborators.  Actually, the first
application of the photomultiplier to scintillation counting was done by
Curran and Baker$^9$.  The work was described in a classified report issued in 1944, but was
only published in the open literature in 1948.  In 1947-1948, Marshall,
Coltman and collaborators published a series of papers
$^{10}$ describing the design and performance of a
photomultiplier scintillation detector, with a well-designed optical
system for reflecting the scintillation emission onto the photocathode.
They reported the detection and counting of $\alpha$-particles,
protons, fast electrons, $\alpha$-rays, $\gamma$-rays and neutrons.
\vskip 12pt
At about the same time, in Germany, Hartmut Kallmann and his student I.
Broser published papers on measurements on scintillations produced by
$\alpha$-particles, $\beta$-rays and $\gamma$-rays in $ZnS, Ca W O_4,
Zn SO_4$ and naphthalene $^{11}$.  That large
transparent blocks of naphthalene, the first organic scintillator and
first large volume scintillator, could produce the photons  from $\beta$-rays and $\gamma$-rays
and subsequently be registered by the photomultiplier tube, represented a
major advance for the new technique.  In 1948, Bell showed that
crystalline anthracene is an even more suitable phosphor and that it
gives scintillation pulses about five times the amplitude of those from
naphthalene. ${^{12}}$
\vskip 12pt
Robert Hofstadter discovered that $NaI$ crystals, activated with
thallium, give higher pulses than anthracene, and because of the high
photoelectric absorption of the heavier iodine constituent, such
crystals can be used for $\gamma$-ray spectroscopy of very weak sources
$^{13}$.
Further developments, using liquid, plastic and crystal scintillators,
soon made the scintillation counter a pre-eminent detector in nuclear
and particle physics.  In recent years, scintillation counting
techniques have found a wide variety of important applications in
biology, chemistry, geology, medicine, atmospheric science, and
industry.
\vskip 12pt
Thus, one can trace the evolution of the modern scintillation counter,
using photomultiplier tubes, from the very humble device produced by
Marietta Blau to its ubiquitous application in all science and
technology.
\vskip 12pt
As a personal footnote, it is interesting to note that the development of scintillation counters
by Robert Hofstadter were critical components of his experiments on the
scattering of (then) high energy electrons (600 MeV) from protons and heavy
nuclei during the 1950's, for which he received a Nobel Prize for
Physics in 1961.  When I was a graduate student under Hartmut Kallmann at
New York University during 1951-1955, several of my colleagues and
postdocs were working on the properties of organic phosphors for
counting $\gamma$-rays and $\beta$-rays (I worked on the
photoconductivity of $ZnS$ and $CdS$ phosphors).  During that period, I
recall several visits by a shy, polite, young man, Robert Hofstadter,
who came from Princeton University to consult with Hartmut Kallmann on
scintillation counters.  A bit later, Hofstadter moved to Stanford
University, where he continued his classic studies of nuclear
structure using the high energy electrons.
\vskip 12pt
It is a further somewhat remarkable confluence of trajectories that
Robert Hofstadter and I became friends when he came frequently during
the 1960's, 1970's and 1980's to the Center for Theoretical Studies at the
University of Miami to visit my colleague, Behram Kursunoglu and me,
and also to participate in a number of the Coral Gables Conferences on
Symmetry Principles and High Energy Physics, and  in
several projects of the Center.
\vskip 12pt
During my years in Kallmann's laboratory, perhaps in 1953 and 1954, I
recall that he received two of those middle-of-the-night phone calls from
journalists in Stockholm, with the news that he was on a list of two or
three finalists to
receive the Nobel Prize for Physics.  Though he was recognized as one of
the principal architects of the scintillation counter, Kallmann shared
the disappointment of not gaining this recognition with Marietta Blau, although for perhaps quite
different reasons.  Actually, no person received the Nobel Prize for the
scintillation counter, a formidable device.
\vskip 12pt
It is unfortunate that Blau did not further pursue the development of the
scintillation counter.  I would venture the explanation that because she
was working for profit-making companies, she was not free to pursue her
own inclinations, but had to follow the directives of management.  In
his biographical sketch, Leopold Halpern relates a conversation with Otto
Frisch, who said that Blau's method of using photomultipliers later
became of great importance $^5$.
\vskip 12pt
\leftline{C. Research on Radioactivity}
\vskip 12pt
During the following two years Marietta Blau carried out a number of
projects involving measurements on radioactivity.  Given that at this
time her employers were mining companies, i.e., International Rare
Metals Refinery, Inc. and Canadian Radium and Uranium Corporation, it is
not surprising that her work involved the studies of devices and
procedures that make use of radioactive substances.  This was certainly
true of her work on the scintillation counter described in the previous
section$^{B46}$.  In that paper she suggested the use of a radium
preparation and the ZnS screen as a secondary standard in calibrating
the intensity of any light source.  One needs to know only the spectral
sensitivity of the photomultiplier (expressed in amperes/lumen), and to
find the efficiency of the fluorescent screen, defined as the ratio:
energy of light emitted by the screen/energy of alpha particles absorbed
by the screen.  She concludes that article by suggesting that the device
is not limited to alpha-or-beta-measurements, but suggested its
application to measurements of strong neutron sources.  However, this
further work, which would surely have led to advances in her
scintillation counter, did not appear, corroborating my surmise that the
corporate leaders chose not to pursue this subject any further.
\vskip 12pt
Her next publication, on ``Radioactive Light Sources'', was co-authored
with I. Feuer, and appeared in the Journal of the Optical
Society$^{B47}$.  In this paper Blau examined further the use of
radioactive preparations to produce light from fluorescent screens.  She
notes that fluorescent screens mixed with radioactive material, such as
those used as paint for watch and instrument dials, change the
luminescent emission because the continual bombardment by alpha
particles changes the crystal structure of the phosphor.  But separation
of radioactive source from the fluorescent screen allows irradiation of
the screen during relatively short intervals, when the light output is
quite constant.  Certain phenomena of fatigue which occur after a
somewhat longer or stronger irradiation are transitory, and the initial
efficiency is restored after a short rest.
\vskip 12pt
The most convenient radioactive preparations -- if absolute constancy is
required -- is radium (half-life 1500 years) placed on a metal foil in
such a manner that there is no escape of emanation and a minimum absorption of
the emitted alpha-particles.  The constants of these radium foils
(number and energy of the alpha-particles emitted) can be tested very
accurately.
The only inconvenience of these preparations is the penetrating gamma radiation,
which may be disturbing in case of radioactive light standards of great
intensity and luminous surface, making necessary the use of stronger
radium preparations.
She suggests the use of polonium instead of radium.  Polonium emits
practically no penetrating radiation, and although its half-life is only
140 days, the intensity of the light source can be calculated exactly
from the exponential decay law.
\vskip 12pt
The paper goes on to discuss the experimental arrangements of
radioactive sources (both radium and polonium) and fluorescent screens,
showing the dependence of light intensity on distance of the sources
from the screens, and on the absorption by aluminum foil.
\vskip 12pt
Blau and Feuer now enumerate a number of  applications and
advantages of this arrangement.  The radioactive light standards could
be very useful for colorimetric measurements and similar purposes.  They
have the advantage of being easier to handle than ordinary light
standards as they do not involve electric currents which must be kept
constant.  Besides, in the case of radioactive light standards, the
light output can be varied by varying the intensity of the radioactive
preparation or the absorption of the radiation, whereas in the case of
light standards, an increase in current or absorption influences the
spectral distribution of the source.
\vskip 12pt
They emphasize that the advantage of radioactive light standards is even
greater on cathode-ray screens (television, oscilloscopes or [now]
computer) or of the luminous compound used for such screens.  They
compare the effects of $\alpha$-particles and cathode rays, and say that
the radioactive source can be directly introduced into the cathode-ray
tube to control and measure the thickness of the screen.  As this can be
done before the tube is closed and evacuated, it might save time and work.
\vskip 12pt
They conclude the article with designs where the radioactive standard is
used to control and regulate the photoelectric section of an x-ray
apparatus, and another where a device is used to control the maintenance
of the level of a bar, disk, or plate.  These are clearly industrial
applications.
\vskip 12pt
The next paper was co-authored during the same year (1946) with H. Sinason
and O. Baudisch, on ``Radioactivation of Colloidal Gamma Ferric Oxide''
in Science$^{B48}$.  This work clearly addresses important issues in
medicine and cancer therapy.
\vskip 12pt
Apparently, the lattice structure of $\gamma~ Fe_2O_3$ is incomplete,
containing so-called ``interstitial spaces'' (atomic holes) which in the
more stable lattice of $\alpha~ Fe_2O_3$, are filled up by ferric ions.
Gamma ferric oxide in colloidal form may be injected directly into the
blood stream.  It had been noted that colloidal $\gamma~ Fe_2O_3$ has no
toxic effects on living cells; the cells proliferate and, after
some time, have eliminated and are entirely free of all the iron
particles.
\vskip 12pt
It is early seen that the reticulo-endothelial cells in the body can be influenced
and their antibody actions stimulated if the colloidal $\gamma~ Fe_2O_3$
particles are combined with some therapeutically acting material, such as
certain radioactive substances.
\vskip 12pt
After rejecting the use of radium and polonium, they finally chose  the active deposits of radon
--$RaA,RaB,RaC$-- as activators.  Because of
their lifetimes, the injections would decay with the lifetime of $Ra B$
(26.8 minutes).  After considering various methods of activation of
metal disks, they adopted the method of Blau's earlier mentor, H.
Petterson$^{14}$.  They conclude that there is no limit as to the charge
of $Ra B - Ra C$ that may be applied with the $\gamma~ Fe_2O_3$.  They
remark that it may be even more advantageous to use, instead of $Ra B -
Ra C$, the active deposit of thorium, Th B and Th C.  Th B has a longer
half-life (10.6 hours), and Th C is the last radioactive element
produced in this series, as Th D is already a stable element.
\vskip 12pt
Finally, they suggest that if we reduce by the radio therapeutic method
the number of circulating lymphocytes, there is some hope to reduce also
the growth and occurrence of tumors.
\vskip 12pt
It is interesting to note that in this work, Blau made an
apparently effortless transition from her early work on medical physics
and polonium preparations$^{B15}$.  I am not competent to judge the
importance of this work for later developments in medical physics, but I
am impressed by its motivation and erudition.
\vskip 12pt
The next paper from this period was co-authored with J. Carlin, on
``Ionization Currents from Extended Alpha-sources'' in the Review of
Scientific Instruments$^{B49}$.  The work studies the ionization
currents from extended two-dimensional radioactive alpha-particle sources,
with immediate application to a device for measuring surface areas.
\vskip 12pt
They first address the problem of the influence of the recombination of
ions on the ionization current produced by the radiation.  They study
carefully the influence of voltages on the electrodes of the ionization
chamber on the saturation ionization current.  In general, the problem
of predicting the saturation current or voltage by theoretical
computations is a cumbersome and complex task, notably because of the
constants involved, which have not yet been precisely determined and are
usually derived from partially limited theories.
\vskip 12pt
By choosing one particular geometrical arrangement of the experimental
apparatus, however, the problems become considerably simplified and the
relationships between saturation current and voltage, if determined
experimentally for one intensity, can be extended by calculations to any
other source intensity.
\vskip 12pt
They describe in some detail the experiments on a large parallel-plate
condenser as the ionization chamber, contained in a cubic housing 30 cm
on each side, with the plates of the chamber each possessing a diameter
of 20 cm.  The insulated electrode which led to an electrometer was
protected by a guard-ring.  The sensitivity of the instrument, a
Compton electrometer, was recorded and followed during the experiments
by measurements with a $U_3O_8$ standard.
\vskip 12pt
The sources employed were circular brass or nickel disks uniformly
plated with polonium (3.83 cm alpha-particle range).  Both the disk area
and polonium density were varied.  The uniformity of plating on each
disk was determined photographically and only the most uniform
preparations, within 5 percent, were studied.  The intensities of the
more active disks were actually measured by the scintillation counter of
Blau and Dreyfus$^{B46}$ described in Part B of this paper.  They go on
to study in great detail how the saturation current and saturation
voltage are influenced by geometrical arrangements, and especially by
the area of the source.
\vskip 12pt
They finally apply their studies to describe a measuring instrument for
the determination of plane areas, which they call a ``polonium
integrator'', where unknown areas interposed between the plates of the
ionization chamber reduce the total ionization current by an amount
proportional to the area.  They actually show a model of the instrument,
which can measure areas from 0 to 28 square inches (175 square cm) with
an error not exceeding one percent.
\vskip 12pt
During the same year, Blau and Sinason published a short paper ``Routine
Analysis of the Alpha Activity of Protactinium Samples'' in {\it
Science}.$^{B50}$  The classical method of measuring alpha activity of
protactinium consists of the following procedure:  Samples containing
protactinium are painted uniformly on metal disks, as thin as possible
in order to provide minimum absorption for the alpha-particles of
protactinium.  The current produced by the alpha-particles is measured
by an electrometer or electroscope and compared with that of a uranium
standard.
\vskip 12pt
The current obtained by the protactinium sample, expressed in e.s.u., is
used for the determination of the number of alpha-particles emitted/sec,
N, $J_{esu} = N\cdot  e\cdot n$, where $e$ is the electric charge of the
ion; and $n$ is the number of ion pairs produced by an alpha along its
path through the ionization chamber.  Knowing the half-life of
protactinium, the number of alpha-particles emitted by, for instance, 1
mg of protactinium can be calculated.  One mg of protactinium emits (in
all directions) $1.85 \times 10^6$ alpha-particles/sec.
\vskip 12pt
The fraction ${2N\over 1.85 \times 10^6}$ gives the amount of
protactinium (in mg) which the sample contains, provided that the
absorption of the layer of foreign material can be neglected.  But this
can never be realized, and the maximum value of the above ratio
corresponding to zero absorption has to be determined by extrapolation,
measuring samples of decreasing total weight.  The ratio (current)/(sample
weight) increases with decreasing layer thickness and, plotting these
values versus sample weights, gives a curve whose intersection with the
ordinate gives the actual value of the protactinium content per unit
weight of the material.
\vskip 12pt
It is evident that this method is time consuming and subject to various
errors, especially if the sample is very dilute.  Moreover, the method
presents a great many difficulties in plants where other radioactive
products are present.
Blau and Sinanson decided to apply for routine measurements the method of
measuring the samples in thick (alpha-saturated) layers, especially
since it affords less handling of the material, which after suitable
grinding, is fitted into special, strictly uncontaminated dishes.
They proceeded by adding a known quantity of polonium to $Zr P_2 O_7$
samples of low protactinium content.  They were then able to calculate
with considerable accuracy the fractional content of protactinium in
their sample.
\vskip 12pt
They recommend the same procedure in the case of other alpha emitters of
appreciable half-life, e.g., plutonium and other transuranic elements,
since the addition of polonium, due to its high specific activity, does
not alter the absorption in the sample.
\vskip 12pt
Finally, in 1948, Blau and Carlin published a paper, ``Industrial
Applications of Radioactivity'' in the journal {\it
Electronics}.$^{B51}$  If I am permitted a somewhat cynical comment,
this paper does not describe new results, but serves as a kind of
announcement of new radioactive devices to the engineering community.
It makes use of the results of Blau and her collaborators over the
previous three years, and cites as references six patent applications of Blau and
collaborators, as well as two other patent applications, presumably also by other employees of
her employers.  Missing from the list of patents is the seminal work of
Blau and Dreyfus$^{B46}$ on the scintillation counter, which I described
in Section B and in the beginning of Section C.
They give technical details of representative new devices based on
radioactive sources,
serving as resistors, electrostatic voltmeters, light sources, tube
cathodes, area measurers, liquid level detectors, galvanometers,
semimicrobalances, leveling systems, and micrometers.
\vskip 12pt
This paper could be looked upon as the swansong of Blau's work for
industrial companies.
\vskip 12pt
In the same year, Blau wrote a paper with J.E. Smith, still as a member
of Canadian and Radium Corp., on ``Beta-ray Measurements and Units'',
which transcended her papers of the previous four years, although they
made use of some of her earlier experiments$^{B53}$.  Here she
attempted to establish a sensible unit to measure the ionizing power of
beta-rays in condensed materials.
\vskip 12pt
The authors discuss the introduction of a new unit proposed especially
for medical research, ``roentgen equivalent physical'' or rep, by
Evans$^{15}$.  This unit later came to be called ``roentgen equivalent
man'' or rem. [The modern SI unit for dose equivalent is ``sievert''
(Sv), where 1Sv = 100 rem].  Blau and Smith do not suggest the
replacement of the roentgen unit as far as x- and gamma radiation are
concerned, but in the case of beta radiation, it would appear that a
more convenient unit is desirable.
\vskip 12pt
They describe the difficulty of making definitive dosage measurements on
beta rays.  While the rep unit indicates the ionization density in the
affected tissue, it does not take into account the amount of tissue
which is actually affected, nor does it give any information about the
total energy administered.
\vskip 12pt
They suggest that a more convenient unit would be a quantity
proportional to the total number of ions formed by the incident
radiation.  They propose a convenient and practical measurement which
can be used by persons not very familiar with radioactive measurements.
They propose the use of a photomultiplier tube and appropriate low
persistence fluorescent screens, emitting light in the range of maximum
sensitivity of the photocell.  This is a clear application of the first
scintillation counter developed by Blau and Dreyfus.$^{B46}$
\vskip 12pt
They go on to describe experiments with beta rays and various
thicknesses of luminescent screens and find good agreement between
theory and experiment.  There is a quite prescient statement where they
propose that the method could be improved by the use of organic phosphors
such as naphthalene, which are more transparent to their fluorescent
light than inorganic phosphors.  Here they quote Coltman and Marshall
$^{10}$, among others since they presumably have not yet seen the
reports of Kallmann and Broser$^{11}$ on naphthalene.
\vskip 12pt
Their method allows them to obtain one of the proposed units, either Q or
Q/E or Q/R, where Q is the total number of ions, E the energy and R the
range of the radiation.
\vskip 12pt
They then describe their apparatus, which is a rather different upgrade of
their first scintillation counter$^{B46}$, and which permits them to examine
ten fluorescent screens with a flick of a dial, and thus record the
photoelectric current, for various measurements.
\vskip 12pt
This paper is quite remarkable both in that it addresses fundamental
questions of radioactive dosage, and proposes the use of a quite novel
instrument to make radioactivity measurements.  It reinforces my earlier
conjecture that Blau made a fundamental contribution to the
scintillation counter.
\vskip 12pt
\leftline{D.  Marietta Blau at Columbia University}
\vskip 12pt
In 1948, the newly established Atomic Energy Commission set up Blau  at
Columbia University as a research physicist, and then two years later
moved her to Brookhaven National Laboratory, which was just then turning
to high energy research.$^1$
\vskip 12pt
Her subsequent research at Columbia and Brookhaven represents a sharp
departure from that of her previous three years for the  mining
companies. Although I have no direct knowledge of the circumstances of
this transition, it is clear from reviewing her papers of the following
period that she exclusively returned to her primary research interest, the use of
photographic emulsions, and their application, to study the phenomena of
high energy physics, with exceptional dedication and energy.  While it is
true that she demonstrated enormous skill and loyalty in her studies of
radioactivity for the mining companies during 1944-1947, she appears to
have thrown herself totally into the emulsion work from which she had
been separated for ten years, while she was effectively deprived of the
recognition she should have had.  Her output during this time is
prodigious, in spite of the fact that she was effectively an outsider in
the American research establishment.
\vskip 12pt
In 1948, M. Blau, then at Columbia University, and J.A. Felice, then at
Brookhaven National Laboratory, published a paper, ``Development of Thick
Emulsions by a Two-Bath Method''$^{B52}$.  Their proposal is alternative
to the method of the so-called temperature development on Ilford, C2,
200 micron plates by Dilworth, Occhialini, and Payne$^{16}$, and is
probably applicable to thicker emulsions.
\vskip 12pt
They adapted a method used by Crabtree$^{17}$ et al, which was used for
the uniform development of large quantities of motion picture film.  The
developer is divided into two baths.  The first part contains the
developing agent part of the sodium sulfite and potassium bromide, but
no alkali.  The second bath contains all the necessary constituents of
an ordinary developer plus an additional amount of alkali.  They then
give an explicit recipe for the two baths.
\vskip 12pt
Because the temperature is kept constant, the danger of reticulation is
avoided.  Proton tracks in the emulsion had their normal grain density
while the background fog was very low, and the plates appeared to be
uniformly developed throughout the emulsion.  This method is discussed
by Galison$^1$, Rotblat$^{18}$, and Powell et al$^7$, and appears to be a
significant contribution to the technique of emulsion development.
\vskip 12pt
Thus, Blau's first foray into the emulsion field, after ten years absence, led
to an important advance, characteristic of her insight and meticulousness.
\vskip 12pt
Blau's next papers $^{B54,B55}$ came out in the following year, and were
on ``Grain Density in Photographic Tracks of Heavy Particles''.  Here she
dealt with one of the three principal parameters of particle track
measurements in emulsion, the other two being the energy-range relations
and the scattering effect, in determining the energy and mass of
nuclear particles.  (It should be noted that grain density is the number
of developed silver bromide grains per length of particle track, and
generally increases as the ionization probability increases.  It plays a
role similar to ionization energy in work with cloud chambers or
proportional counters.)
\vskip 12pt
She refers first to an empirical relation between grain density and
range which was established by workers in Powell's laboratory and others
$^{19,20,21,22}$.  She then describes the development of
theoretical formulas for the dependence of grain density on range and
energy, based on the fundamental theory of Debye and H\"uckel$^{23}$,
who solved the problem of highly ionized gaseous atmospheres or
``strong electrolytes'', by assuming that the mobility of ions depends
on the ion concentration.  Comparing her formulas with the empirical
results of the Bristol group$^{19,20,21}$, she found very good
agreement.
\vskip 12pt
It is quite striking that in this paper, and in the previous one on
emulsion development, Blau methodically reeducated herself on the
techniques of experimental emulsion work and acquainted herself with the
principal technical developments since 1938.  In both papers she made
significant contributions to the field.
\vskip 12pt
Blau's following paper was done with M.M. Block and J.E. Nafe$^{B56}$,
on ``Heavy Particles in Cosmic Ray Stars''.  This signalled the
beginning of her particle studies at Columbia University, where she was
apparently brought in to instruct the researchers on the techniques of
using photographic emulsions at the Nevis cyclotron, which was then
under construction.  Apparently, for practice, they exposed some
emulsions in balloons, and came upon a strange event in a cosmic ray
star$^{24}$.  In those years, as the ``elementary particle zoo'' was
being assembled, each event was afforded special attention.  In this
case, the event was interpreted as the capture of a $\tau$-meson (now
called a K-meson or kaon) by a bromine or silver nucleus in the
emulsion.
\vskip 12pt
The next paper, by Blau, Ruderman, and Czechowski, on ``Photographic
Methods of Measuring Slow Neutron Intensities'' appeared in
1950$^{B57}$.  The relative slow neutron sensitivities of
$\beta$-sensitive emulsions, x-ray film-indium foil combinations, and
boron-loaded plates were investigated.  Since the detection efficiency
depends upon the neutron energy, experiments were made with epithermal
(0.3-10,000eV), thermal (.01 - 0.3eV) and cold ($< .01$eV) neutrons.
$\beta$-sensitive emulsions and x-ray-indium combinations are about
equally useful for the detection of epithermal neutrons.  $B^{10}$
loaded plates, which are best for detection of thermal and cold
neutrons, have the following advantages:  very low neutron intensities
can be measured by counting of $\alpha$-tracks, neutrons can be counted
in the presences of $\beta$- and $\gamma$-radiation, the number of
$\alpha$-tracks is independent of the development conditions, and a wide
range of intensities can be measured with a single plate.
\vskip 12pt
Blau's next publications,
during the same year, were on a ``Semi-Automatic
Device for Analyzing Events in Nuclear Emulsions'' in the Physical
Review , with S. Lindenbaum and R. Rudin $^{B59,B62}$.  This was a
landmark work that led not only to future advances in analyzing emulsion
tracks, but also portended much later developments in the analysis of bubble
chamber, spark chamber, and streamer chamber photographs.  The
contributions of Marietta Blau to the Nevis Cyclotron experiments and to
the semi-automatic device described in this paper are discussed cogently
by Sam Lindenbaum in Appendix II of this paper$^{25}$.  I must confess the sin
of indolence in that I have not scoured the literature for references to
this work in later developments of automatic devices for scanning and
measuring visual tracks in detectors.  (I am not a dedicated historian
of science).  When one considers the rudimentary level of computers and
optical devices four decades ago, the capabilities of this device are
remarkable, and deserving of admiration.
\vskip 12pt
The instrument is built around a microscope with a motor-driven stage,
moved by selsyn motors in two dimensions.  The accuracy of gears,
feedscrews etc., is such that dimensions can be measured to within 0.2
micron.  These selsyn motors are fed from identical selsyn generators
which are driven from a steering unit so that the photographic plate can
be moved in any direction and with any desired speed up to 25 microns
per second.  For convenience, the arrangement is controlled by a steering
wheel and the speed is controlled by a foot pedal.  There is also a
recording chart that moves at a speed of 2000 times that of the stage.
The image of the plate is observed by the operator through the eyepiece
and at the same time is projected on a small slit before a
photomultiplier tube.
\vskip 12pt
The measurements are done quite rapidly and easily.  For example, a
conservative estimate of the driving time for the grain density record
of a
2000 micron track is about 10 minutes.  The system is adaptable to the
measurement of high Z tracks (3$\leq$Z$\leq$26), especially by the
measurements of $\delta$-rays (electron tracks emanating from the main
track).  They compare their measurements with those of Bradt and Peters
on large Z nuclei and P. Freier$^{26}$, and find good agreement.  The
paper is notable for its completeness and erudition.
\vskip 12pt
A brief paper, ``Dependence of High Altitude Star and Meson Production
Rates on Absorbers'' written by Blau, Nafe and Bramson was given as an
abstract in the Physical Review in 1950$^{B60}$.  They studied the
effects of copper and lead absorbers on the rate of star and meson
production.  Comparing star and meson production in the different sets
of plates under varying thickness of absorbers, they observed distinct
transition effects.
\vskip 12pt
\leftline{E.  Marietta Blau at Brookhaven National Laboratory}
\vskip 12pt
After her move to Brookhaven National Laboratory in 1950, Marietta Blau
published her first paper, an abstract, ``Stars Induced by High Energy
Neutrons in the Light Elements of the Photographic Emulsions'' in the
1952 Physical Review $^{B63}$, with A.R. Oliver.  They irradiated
emulsions with the 300 MeV neutron beam of the Columbia cyclotron and
determined the ratio of stars induced in the light elements to those
induced in the heavy elements of the emulsion.  Assuming the cross
section is a linear function of $A^{2\over 3}$ (where $A$ is the atomic
mass), the ratio N-light/N-heavy should be 0.27.  Counting all stars
with $\geq 2$ prongs in both emulsion and gelatin layers, they obtain a
value for this ratio of 0.179$\pm$.024.  Because of the uncertainty
connected with the recognition of 2-prong stars only 2-prong stars
showing a distinct recoil fragment were considered.  On the other hand,
they accepted all 2-prong gelatin stars assuming that in all cases an
additional short prong may have been lost in the gelatin.  Even though
these conditions favor a higher ratio N-light/N-heavy = 0.213$\pm$0.026
it is still lower than the calculated value, showing increased transparency
of light nuclei.  If one takes into account the 0-, 1-, and 2-prong
stars$^{27}$, N-light/N-heavy would still be further decreased since for
the light elements 3- and 4- prong stars are probably favored.
\vskip 12pt
In the following year, 1953, Blau, Oliver and Smith expanded on the
previous abstract with a longer paper on ``Neutron and meson stars
induced in the light elements of the emulsion'' in the Physical
Review$^{B65}$.
\vskip 12pt
Again, they made use of G5 emulsions, laminated with gelatin layers of
5-8 microns.  They followed the ideas of Harding$^{28}$, Menon, Muirhead
and Rochat,$^{29}$ and Hodgson$^{30}$ of introducing very thin layers of
gelatin between photographic emulsion pellicles in order to separate the light
and heavy emulsion elements in experiments investigating the cross
section of disintegration processes of particles incident on the emulsion.
\vskip 12pt
They expand somewhat on the results of the previous paper on 300 MeV
neutrons, finding that the mean prong number in stars from light nuclei is
greater than in heavy nuclei.  The $\alpha/p$ ratio for light nuclei is
approximately  0.75, and 25$\pm$5 percent of all stars have a recoil
with charge $\geq 3$.  From the angular distribution of black tracks in
light element stars, and the forward excess of black prongs in heavy
elements, they conclude that, at most, 70 percent of black prongs in
heavy elements are due to nuclear evaporation.
\vskip 12pt
The second part of the paper discusses the results of exposing the same
configuration of emulsion pellicles and gelatin layers to the positive pion
($\pi^+$-mesons) beams of 70-80 MeV and 60$\pm$5 MeV from the Columbia
cyclotron.  They conclude that 24-30 percent of emulsion stars originate
in the light nuclei of the emulsion, and find a lower limit of the
opacity of light nuclei of 0.64.  In most cases, absorption of the
incoming meson takes place, and the absorption occurs mainly on nucleon
pairs.  Absorption by more than two nucleons is less than 30 percent.
\vskip 12pt
In an interesting Appendix to this article, the authors describe the
absorption of x-rays by the emulsions used in these experiments.  They
finally determined that the ratio of light elements in emulsion to light
elements in the gelatin layers is $2.95 \pm 0.3$.
\vskip 12pt
In 1952, Blau and Salant published a letter in the Physical
Review$^{B64}$ on ``T-tracks in Nuclear Emulsions.''  This paper
presents evidence of so called ``T-tracks'', which seemed to be heavy
particles produced in the cosmic radiation and stopping in the emulsion,
with the apparent release of other particles, including fast
(minimum-ionizing) particles.  The eleven cases they observed seem to
represent some kind of enigma.  They cannot decide whether the T-tracks
and their products are coincidences of observation or not.  Since I have
not seen any further reference to this phenomenon, I would have to
conclude that the observations were not valid.  I do recollect that Blau
was not fond of Salant, and that he may have been a cause for her
eventually leaving Brookhaven.
\vskip 12pt
The next two papers were produced from work at Brookhaven National
Laboratory on the interactions of 500 MeV negative pions
($\pi^-$-mesons) produced with emulsion nuclei.  The first was a letter
with M. Caulton and J.E. Smith in Physical Review in 1953$^{B66}$ and
then a
longer article with M. Caulton in the Physical Review in 1954$^{B67}$.
These papers signalled the complete return of Blau to nuclear physics
research with the technique she had essentially invented, and in which she
kept pace with the widening application of photographic emulsions to
high energy subnuclear phenomena.  According to Galison$^1$, she was the
first to demonstrate that meson interactions with nuclei could produce
additional mesons$^{B66}$.  Actually, in that letter, Blau, Caulton and
Smith give credit to the Bristol group$^{31}$ for finding a meson
production event in plates exposed to cosmic rays and one event found in
the 220 MeV negative pion beam of the Chicago cyclotron$^{32}$.
However, since the authors identified six events in which two mesons
leave an emulsion nucleus and a seventh in which two mesons appear to
emerge from a hydrogen nucleus (a proton), it gives unequivocal evidence
of meson production.  In addition to the six events, eight more are
consistent with additional meson production, although the tracks are too
short for unmistakable identification.
\vskip 12pt
Hence, it may be stated with considerable justification, that Blau's was
the first definitive report of additional meson production by high
energy mesons, an important, if not unexpected, observation.
\vskip 12pt
They estimate that the number of two meson events reported in the paper
constitute at most 25 percent of all events in which a charged meson -- observed
or absorbed -- is produced.  The above figure would then represent 5.2
percent to 12 percent of all interactions.  From this experiment nothing
can be learned about the production of neutral mesons
($\pi^0$-mesons) and therefore the actual cross section of
meson-meson production at 500 MeV meson energy cannot be compared with
theoretical data.  They quote, however, C.N. Yang and E. Fermi (a private
communication), who estimate that the fraction of events leading to two
charged mesons is 16 percent to 18 percent (the second number includes
events with two charged and one neutral meson).  This figure is twice
the value calculated by Blau, Caulton and Smith from the experimental
data, but considering the meager statistics and the simplified
assumptions, the disagreement is probably not too serious.
\vskip 12pt
In this paper, I can recognize the total modesty, honesty, and
commitment of Marietta Blau, who ventured no claims beyond the
experimental evidence and conservative conjecture.
\vskip 12pt
The second, longer paper, ``Inelastic Scattering of 500-MeV Negative
Pions in Emulsion Nuclei''$^{B67}$ was on an expansion of the results
given in the shorter letter $^{B66}$.  It goes into considerable detail
on the exposure of the emulsions to two different beams of the
Brookhaven Cosmotron, namely (1) the 500 MeV meson beam and (2)
particles emitted from the Cosmotron target (beryllium) at an angle of
$32^{\circ}$ with the 3 GeV proton beam direction.  They also describe the
methods on the search for interactions in the pellicles, and on the
measurements made on incoming meson tracks and on outgoing particles.
\vskip 12pt
They discuss, in succession and in some detail, the observations of stars
with no mesons, stars with one meson and events with two mesons, i.e.,
additional meson production.  They perform a careful analysis of meson
interactions in nuclear matter, and also some data on the production of
neutral pions ($\pi^0$-meson).  They finally conclude that the
cross section for the production of charged mesons per nucleon is 3.5
millibarns or 14 percent of the total meson -- nucleon cross section at
500 MeV.  They believe that they underestimate this cross section, and
that it could be as high as 10 millibarns, probably an overestimate.
They note that C.N. Yang and E. Fermi (a private communication) expect
the cross section for meson production by mesons to be 12 percent of the
total cross section, which is estimated by Lindenbaum and Yuan$^{33}$
to be 610 millibarns.
\vskip 12pt
They conclude the paper with a comparison of the results with those of
cosmic ray mesons, where the mean shower energy is 640 MeV$^{34}$.  There
appears to be a considerable discrepancy with the results between shower
mesons and single artificial mesons, which they are at a loss to
explain.
\vskip 12pt
Blau resumed her collaboration with A.R. Oliver and continued her
studies of high energy pion interactions, publishing a paper on
``Interaction of 750-MeV $\pi^-$-mesons with Emulsion Nuclei'' in the
Physical Review$^{B68}$ in 1956.
\vskip 12pt
The pions were selected by an analyzing magnet from secondary particles
emitted from a beryllium target at 32$^{\circ}$ to the direction of the
proton beam of the Brookhaven Cosmotron.  They scanned 132.8 meter of
meson track under high magnification and found 322 interaction events.
Subtracting from the path length ($6.5 \pm .02$)\% for probable muon
contamination, the mean free path in emulsion is ($38.5 \pm 2.2$)cm, while
the geometric mean free path (for $r_0 = 1.38 \times
10^{-13}cm)$ is 27 cm.  This is in fair agreement with a value expected
from the $\pi^-$ - $p$ cross section of 42 millibarns and $\pi^-$ - $n$ cross
section of 17.5 millibarns found by Lindenbaum and Yuan$^{33}$ in the
same energy interval.  Accepting these cross sections and integrating
over the emulsion nuclei (number of neutrons = $1.2 \times $number of
protons) leads to a mean free path of 34.7 cm.
\vskip 12pt
They then analyze the results according to the type of interaction.  They
find 5 elastic scattering events on free protons and 6 near elastic
scatterings on protons near the edge of nuclei.  From the 5 events, they
find a mean free path for elastic $\pi^- + p$ interactions of $24.8 \pm
11.4$ meters, while the mean free path for all $\pi^- +p$ interactions
(cross section 42 millibarns) is expected to be 7 meters; the elastic
cross section without charge exchange is $12 \pm 5.4$ millibarns or
about one third of the total cross section.  If one assumes the relation
$$
(\pi^- + p \longrightarrow \pi^- + p)/( \pi^- + p \longrightarrow n +
\pi^0) =
{5\over 4}~({\rm equal~weight~for~states~}{3\over 2}~{\rm and}~{1\over
2}),
$$
\noindent
then about 60 percent of all interactions should be elastic.
\vskip 12pt
They then turn their attention to 15 events that can be called inelastic
scattering by free protons or protons at the periphery of the nucleus,
according to the schemes
$$
\pi^- + p \longrightarrow \pi^- + \pi^+ + n~{\rm or}~ \pi^- + p
\longrightarrow \pi^- + \pi^0 + p.
$$
\noindent
They show the angular distributions (in the center of mass system), of
emitted nucleons, fast mesons and slow mesons.  Nucleons in the backward
direction is preferred (12:3), and for fast mesons the forward direction
is preferred.  They calculate the Q-value for the nucleon and the slower
meson, assuming they form an ``excited state,'' and find Q-values spread
over an interval of 30-170 MeV, and there is, at least within small
statistics, no indication of a maximum value.
\vskip 12pt
In the study of charge-exchange scattering and a possible $\pi^- + p
\longrightarrow \pi^0 + \pi^0 + n$ reaction, they find two
events in which electron pairs (Dalitz pairs) give direct evidence for
$\pi^0$ production.  They find 15 events where the meson stops in
the emulsion and another 16 where there is a small recoil or blob.  The
number of stoppings observed (15), plus the two cases with electron
pairs, is unexpectedly high in comparison in the cases leading to
charged meson production, even within the meager statistics$^{35}$.

\vskip 12pt
Finally, they find fifteen events of meson scattering through angles
$\geq$ 10 degrees without visible nuclear interaction and 17 other cases
where the meson scattering is accompanied by recoil or slow-electron
emission.  The first 15 events could represent elastic scattering on
peripheral neutrons or scattering with $\pi^{\circ}$ production; but some
of these could be scattering on protons in nuclei where the emission of
the slow proton has been suppressed by the exclusion principle.  There
is one case observed of $\pi^- + n \longrightarrow p + \pi^- + \pi^-$.
\vskip 12pt
The total number of $\pi^- + p$ collisions  is 41, taking 11 as the number
of elastic scatterings observed.  The mean free path in the emulsion is
$3\pm 0.5$ meter or about one-half of the mean free path for $\pi^-$
collisions on free protons, calculated with a cross section of 42
millibarns.  Therefore, one-half of all observed collisions must have
occurred on bound protons on the nuclear periphery.
\vskip 12pt
The find the ratio
$$
{{\pi^- + p \longrightarrow \pi^- + p}\over {(\pi^- + p \rightarrow
\pi^0 + \pi^- + p) + (\pi^- + p \rightarrow \pi^+ + \pi^- + n)}} =
{11\over 13}.
$$
\noindent
Judging from this ratio and assuming again equal weights for the 3/2 and
1/2 state in Fermi's theory, 44\% of all $\pi^- + p$ collisions are
inelastic.
\vskip 12pt
Finally, the ratio of $\pi^- + p$ to $\pi^- + n$ scattering (considering
in both cases events without evaporation tracks or recoils) is 41:16=2.6,
while the ratio found in this energy interval by Lindenbaum and
Yuan$^{36}$ is 42:17.5=2.4.
\vskip 12pt
They then turn their attention to interactions of the $\pi^-$-mesons
with nuclei.  In addition to 266 nuclear events (233 stars, 17 meson
scatterings with recoils or electron interactions), they found also 500
stars in area scanning.  They find $(40 \pm 4)$\% of all stars have 1
emitted meson and only $(3\pm 1)$\% have 2 emitted mesons, quite
similar to the results obtained for $500MeV$$\pi^-$-mesons$^{B67}$.  They
explain this by the higher interaction cross-section of 750MeV mesons
and by an increased meson production leading to mesons in an energy
interval of 100-300 MeV which have small mean free path in matter.  They
also discuss the angular distribution of the mesons in the laboratory
system.
\vskip 12pt
Noting that if the $\pi$- meson is completely absorbed, the total
excitation energy is nearly 900 MeV; and in heavy elements (which are
responsible  for all stars), stars with 9-15 prongs would be expected.
On the other hand, it can be anticipated that stars with fast
forward-scattered mesons are small, consisting of a few short prongs in
light elements and 1-2 prongs in heavy elements.  As already mentioned,
60\% all stars have no charged meson; 62\% of these have no fast proton
$\geq 60$MeV, and a mean prong number of only $4\pm 0.4$. Only 2\% have a prong
number $\geq 9$.  This suggests that a
great amount of energy is carried away by fast neutral particles for
which the nucleus is rather transparent.  Finally, the great number of
observed $\pi^-$ stopping raises the question of a possibly greater
proportion of collisions leading to charge exchange than is calculated
with the Fermi statistical theory.
\vskip 12pt
They conclude the paper with a description of prong distribution in all
types of stars, and the appearance of stable fragments in $\pi^-$- meson
stars.

\vskip 12pt
Blau's last work published from research at Brookhaven National
Laboratory was on ``Hyperfragments and Slow K-mesons in Stars Produced
by 3-BeV Protons'' in the Physical Review$^{B69}$.  This work was done
about a year after the first example of a hyperfragment (a
$\Lambda^0$ hyperon bound to an ordinary nucleus) was found by
Danysz and Pniewsky$^{37}$,  and then shortly after the first systematic
investigation of hyperfragments produced by the 3 GeV proton beam at the
Brookhaven Cosmotron by Fry, Schneps and Swami.$^{37}$  A systematic search for hyperfragments in
particle beams of well-defined energy gives information on data related
to particle physics, $\Lambda^0$-production cross section, particle
nature of the $\Lambda^0$ (associated production, etc.), as well as
to physics of the nucleus (formation of the hyperfragment, binding
energy, etc.).
\vskip 12pt
The stack of emulsions was exposed to about 30,000 protons/cm$^2$ and
then the individual emulsions were scanned at fairly low (300X)
magnification, since she was searching for fairly prominent stars.
She describes in some detail the measurement procedures followed for the
determination of mass, charge, and energy determination of the
hyperfragments themselves and of the outgoing decay particles or
nuclear fragments.
\vskip 12pt
Of the 14,480 stars investigated, she found 14 events which are believed
to be spontaneous disintegrations of  hyperfragments coming to rest in
the emulsion, with the possible exception of one event where decay in
flight is suspected.  In addition, she found two stars with double
centers which may represent disintegration of slow and probably heavy
hyperfragments.  Both cases occurred in large stars and it was impossible
to disentangle the prongs belonging to each center.
\vskip 12pt
Only a few of the events could be analyzed; however all of them, with
one exception are compatible with a $\Lambda^0$ hyperon bound to
the nucleus.  All the observed hypernuclei, with one exception, are
isotopic spin singlets, I = 0 for nuclei with odd atomic number, and
doublets, I = ${1\over 2}$ for nuclei with even atomic number. The
exception could be explained as a decay in flight of a $_{\Lambda}Li^8$, the first
decay of a hypernucleus in a I=1 state.
\vskip 12pt
She goes on to discuss the likely identity of the other hypernuclei,
$_{\Lambda}Be^8$ and $_{\Lambda}Be^9$, a $_{\Lambda}H^4$,  a
$_{\Lambda}Li^6$, $_{\Lambda}B^9 - _{\Lambda}B^{10}$ or
$_{\Lambda}C^{10} - _{\Lambda}C^{13}$,
$_{\Lambda}B^9$, or $_{\Lambda}C^{11} - _{\Lambda}C^{12}$, a
$_{\Lambda}O^{16}$ or higher atomic number, $_{\Lambda}Be^9$,
$_{\Lambda}H^4$ or $_{\Lambda}H^3$.
\vskip 12pt
The number of hyperfragments per star is $14/14,480 \approx 1\times
10^{-3}$, in good agreement with Fry's results$^{37}$, who found 21
hyperfragments
in 20,000 stars.  Since all hyperfragments originate in heavy elements,
which are responsible for only about 75\% of all stars, the frequency is
actually somewhat higher, about $1.3 \times 10^{-3}$.
\vskip 12pt
Blau also has found four $K^{\pm}$ mesons emitted from stars, all of them of
low energy, since the probability of finding particles of range
$\geq$2.5 cm is small in a stack of the size used in the experiment.
She finds no examples of associated production of other unstable
particles.
\vskip 12pt
In summary, from Blau's work at Brookhaven National Laboratory, it is clear
that Marietta Blau stepped authoritatively into the main stream of
particle research, in spite of the fact that she was not in command of
large research groups.  In particular, she quantified the interaction
of (then) high energy pion interactions including finding the first
examples of additional pion production.  She also contributed
significantly to the observations of hyperfragments at an early stage.
Although the improvement of statistics came several years later with the
exploration of hydrogen, deuterium and helium bubble chambers, the path
of further research was made clear by the emulsion results.
\vskip 12pt
\leftline{F.  Marietta Blau at the University of Miami (Coral Gables)}
\vskip 6pt
\itemitem{i)} A Personal Memoir
\vskip 12pt
To the best of my recollection, Marietta Blau came to Coral Gables as an
Associate Professor in Autumn, 1956.  I had arrived in February, 1956 to
take up the position of Assistant Professor, and continued my interests
in solid state physics by then investigating the optical properties of the
semiconductor GaAs.
\vskip 12pt
Because I do not have a very good memory for detail, I cannot recount
the specifics of my meeting with Blau, but I do recall that we had a
nearly instant rapport, and that I responded without hesitation to her
invitation to collaborate with her on photographic emulsion research in
high energy physics.  She also recruited three other colleagues,
including Claude F. Carter.  The other two did not remain with the group.
\vskip 12pt
I feel that I must depart here, at least temporarily from simply
summarizing her papers, and interject my own impressions and experiences.
It was not simply on a whim that I answered her call, because in the
previous few years, before and after my Ph.D. in 1955 in solid state
physics, I had seriously attempted to independently study theoretical
physics, particularly nuclear physics, with some guidance from a former
teacher at New York University. Before I made any substantial progress
on my own, the idea of an opportunity to do high energy experimental
physics with such a distinguished mentor excited my imagination.
\vskip 12pt
Marietta Blau was a rather small person, perhaps 5 feet 2 inches tall
(158 cm) and quite slender, with a sweet kindly expression.  Her head
was barely visible over the steering wheel of her little Plymouth coupe,
and she was not a very skilled driver, yet she negotiated the trip from
New York to Miami several times, before the days of interstate roads.  The initial
impression she made was that of a fragile person who could be blown over
by a breeze.  I would say that she was quite good looking, but presented
herself in a very modest, self-effacing manner.  She spoke deliberately,
slowly, and softly, and her English, if slightly accented, was polished.
She was well-versed in the classics, literature, and the arts.  We
attended many musical events together, especially visiting chamber music
groups.  I recall that we once attended a presentation of Verdi's
``Requiem'', by the Miami Symphony (then the University of Miami
Symphony), and that we were so overwhelmed, that we jointly sent a warm
letter of appreciation to the conductor, John Bitter (a brother of the
well-known authority on high-field magnets, Francis Bitter).  It was my
first hearing of the ``Requiem''.
\vskip 12pt
When she arrived in Miami, she found several former students and
colleagues from Vienna, namely Fritz Koczy and Elizabeth Rona, both of
them at the Institute of Marine Sciences at the University of Miami.
Koczy later became Director of the Marine Institute, and he and I became
great friends until his untimely death, I believe in the 1970's.
\vskip 12pt
My oldest son, Bernard, was three years old when we arrived in Miami,
and Joseph was born in 1957, just after Marietta arrived.  She became a
close family member, showering the children and my first wife, Ruth,
with gifts.  My older son reminds me that she presented them with a
sleep-out Indian tent when he was about six years old.  We often
entertained each other in our homes.  She met my parents on their visits
to Miami, and they all enjoyed each other greatly.  She grieved
compassionately when my mother  died in 1960, shortly after Marietta's
return to Vienna.
\vskip 12pt
I cannot recall the details of how she built up our laboratory, but she
did have generous funding from the Air Force Office of Scientific
Research (AFOSR). Actually, Claude Carter assisted her ably in fiscal and procurement affairs, in which I played just a
peripheral role.  We obtained about six or seven precision Leitz
binocular microscopes with magnifications up to 2000X, for which we had
designed enlarged movable stages, built for us by local instrument
makers.
\vskip 12pt
Initially, we were housed in cramped quarters of the Physics Department,
at that time in wooden shacks which were used during World War II for the
training of Air Force pilots.  The University of Miami then rented space
for us on the ground floor of an old apartment building in Coral Gables,
about two miles away from the Main Campus.  The upper floor was occupied
by the Institute for Molecular Evolution directed by Sidney W. Fox, a
distinguished biochemist from Florida State University.  We had about
eight or ten rooms, mostly devoted to scanning microscopes, equipment
room, and of course the ubiquitous coffee lounge.  One large room was
reserved for a multiple scattering microscope, whose large foundation
and massive stage were designed by a master instrument maker, Brouwer,
from a town near Berkeley, California.  Brouwer was famous for
designing and building these instruments for emulsion laboratories
around the world, for the purpose of measuring the momentum of fast,
minimum ionizing tracks in the emulsion, and an essential parameter in the
determination of particle properties.  I cannot recall the exact
specifications of this marvelous stage, but I seem to recall that it
could measure deviations of a micrometer in longitudinal movements of
several centimeters (some tens of thousands micrometers).  It was
the massive base of concrete blocks for this microscope that required
our laboratory to be placed on the ground floor of the building.  An
interesting dividend of our needs (the microscopes and the fragile
emulsions) was that the entire facility had to be air conditioned (at
that time by window coolers), an unusual luxury in those years in
Miami, which suffers sweltering temperatures for half of the year .
\vskip 12pt
Marietta Blau was a most effective teacher, giving courses in
electromagnetism and nuclear physics, among others, to advanced
undergraduates and graduate students.  At that time, we did not have a
Ph.D. program, but gave a Master of Science degree.  I believe, that
because of her slight stature and her gender, she was not afforded the
respect to which she was entitled.  She fought with the administration
and the Chairman of the Physics Department about the use of overhead
from her federal grants.  After one complaint to another colleague, the
theorist Behram Kursunoglu, I recall that the latter called the chairman
an idiot.  She answered, in her soft plaintive voice, ``Is that all he
is?''.
\vskip 12pt
I really cannot recall how I learned the emulsion craft, but I do know
that Marietta Blau was a wonderful teacher in the laboratory.  We
recruited housewives and students, in the tradition of Cecil Powell, to
be scanners of the emulsion pellicles, and in the cases of the more
gifted assistants, to allow them to make precision measurements.  She of
course schooled me in the theory of ionization measurements, multiple
scattering, and range-energy relations, and I had to study these
problems on my own, but after about three years I emerged as fairly
competent researcher.  All of the faculty scanned, supervised the
scanners, and made precision measurements as needed.  The tedium of
emulsion work cannot be overestimated.  Constant checking of the
efficiency and accuracy of observations was necessary.  It is not a
simple matter
to describe the amount of work that went into the experiments discussed
in Parts C and D of this memoir, nor that which is described below.
\vskip 12pt
As the bubble chamber, spark chamber, streamer chamber, and digitized
chambers evolved in the 1960's and 1970's, emulsion research regressed
in importance in the accumulation of statistics.  One of the reasons is
that the bubble chambers and spark chambers could be composed of
elemental substances, such as hydrogen, deuterium, helium, or other
interesting substances, and that the measurements could be automated to
the extent that made the accumulation of data far more rapid and
efficient.  In modern applications, electronics, solid state detectors,
and other devices further increase the effectiveness of detectors.
However, emulsions remain an interesting detector in special situations,
especially where spatial resolution and continual exposure are needed.
\vskip 12pt
\leftline{(ii) Marietta Blau's Research at Miami}
\vskip 12pt
Sometime during 1957, when the laboratory was established, we began to
scan a small stack of Ilford G5 emulsions which had been exposed by Gus
Zorn to the negative pion beam at the Brookhaven Cosmotron.  The results
were reported in Il Nuovo Cimento, in 1959, by M. Blau, C.F. Carter and
A. Perlmutter, ``Negative Pion Interactions at 1.3 GeV/c''$^{B69}$.
\vskip 12pt
Since the energy of the incident pions was not precisely known, the
momentum was determined by multiple scattering measurements to be $(1.3
\pm 0.1)GeV/c$, which was corroborated by kinematic considerations of
elastic scattering on protons.  The plates were scanned by following
incident pions from the lead edge.  The track length scanned was
380 meters, and the number of interactions found was 811.  After
correcting for scanning efficiency and for the contamination of the beam by muons and electrons, the corrected
track length was 340 meters, giving a total cross section in agreement,
within statistical limits, with the results obtained by other
investigators.
\vskip 12pt
There are tables which summarize the interactions with free or nuclear
edge nucleons.  There are 10 elastic collisions with free protons,
$\pi^- + p \rightarrow \pi^- + p$, and (7) with peripheral protons
(matched by Fermi momentum) (type I); 11 inelastic collisions with free
protons, $\pi^- + p \rightarrow \pi^- + p + \pi^{\circ}$ and (15) with
peripheral protons (type II); and 5 inelastic collisions, $\pi^- + p
\rightarrow n + \pi^+ + \pi^-$, and (1) with peripheral protons (type
III).  Other collisions with peripheral protons which could not be
analyzed total (38).  As for collisions with peripheral neutrons there
are (8) quasi-elastic ones ($\pi^- + n \rightarrow \pi^- + n)$ (type
VIII),
another (29), elastic or inelastic which could not be analyzed, and
another handful of inelastic events.  A comparison of these events with
those of Walker and Crussard$^{35}$ at a somewhat higher energy does not
indicate any significant disagreement.
\vskip 12pt
It is of interest, in view of the results to be presented later, to
substantiate that the clean $(\pi^- + p)$ events of types I, II and III
represent free proton interactions; this supposition could be verified
in part by calculating the total free proton cross section.  But to do
so one needs to know the relative contributions of multiple pion
production and events of type VI with zero prongs $(\pi^- + p
\rightarrow n + \pi^{\circ}~{\rm or}~ \pi^- + p \rightarrow n +
\pi^{\circ} + \pi^{\circ})$ to the total free proton cross section.
Although these contributions could not be found directly from this data,
they could be estimated from the hydrogen cloud chamber results of Eisberg et
al$^{38}$ and from the curves derived from bubble chamber experiments of
the Wisconsin group$^{39}$. Both groups find a total cross section of 34
millibarns.
\vskip 12pt
The cloud chamber results$^{38}$ seem to indicate that one should expect
double production events to amount to about 25\% of those of types (II
and III) and that the number of zero prong stars should be about 15\% of
all visible proton events, including double production.  Applying these
corrections to our data, we obtained a mean free path for free proton
collisions of 9.7 meters (30 millibarns).  The contribution of zero prong
stars could also be estimated from the bubble chamber data$^{39}$,
yielding a charge exchange cross section (elastic and inelastic) at 1.2
GeV of 30\% of the visible proton interactions.  This leads to a mean
free path of 8.7 meters (34 millibarns) for this emulsion data.
\vskip 12pt
Although the size of this sample is small, if it is combined with the
cloud chamber data$^{38}$ at a similar energy, some of the features seem
to suggest a regularity which warrants further investigations on much
larger samples.  The combined results (for types II and III) are plotted
as function of nucleon momentum in the center of mass system, with a
pronounced peak at about 550 MeV/c, in qualitative agreement with both the
statistical$^{40}$ and isobar models $^{41}$ of pion production, and as
function of their angle in the center of mass system, with the
distribution strongly peaked in backward direction, in agreement with
the isobar model$^{41}$.  In both cases it is evident that events which
require Fermi momentum to be added to the target protons cause an
apparent broadening of the distributions, as might be expected.  Further
figures give momentum distribution in the center of mass system of
protons from reaction of type II plus similar data from the cloud
chamber experiment$^{38}$, and the angular distribution of the same protons in the center of mass system.  Similar
plots are given for the neutrons of reaction of type III.  Further plots
of the momentum distribution and angular distribution in the center of
mass system of both emergent pions in reaction II, and nuclear
interactions are given.
\vskip 12pt
Finally, a histogram of the Q-values in the rest frame of the
nucleon-pion isobar for reactions II and III, is given.  Although the
sample is not large, the Q-value does show a bunching near 150 MeV,
while the range of values is from 0 to 300 MeV$^{42}$.
\vskip 12pt
The remainder of the paper discusses the nuclear interactions of the
pions.  the total number of interactions with nuclei is 725 (not free
proton collisions).  In addition to 263 stars without emitted charged
mesons, we find 41\% of all events have no emitted mesons.  The number
of 1 meson: 2 meson: 3 meson stars is 317:94:14.  In plotting the
angular distribution and energy distribution of emitted mesons, a
comparison is made with the Monte Carlo calculations in Los Alamos by
the MANIAC$^{43}$ and shows only fair agreement.
\vskip 12pt
The paper concludes with a listing of events producing $K$ mesons,
hyperons and hyperfragments, a total of ten events.
\vskip 12pt
During Blau's stay in Miami she vigorously attacked the problems of
ionization in nuclear emulsions, the results of which were given in an
article in The Review of Scientific Instruments in 1960$^{B72}$.
This was a problem which she had addressed earlier in articles at
Columbia University$^{B54,B55}$ and which justifiably obsessed her.
During the course of this investigation, in which measurements were made
on known tracks for the purpose of particle investigation, it was found
that a single parameter, namely mean blob length, could be related to
the probability of ionization over the entire energy range.
\vskip 12pt
We discussed first the probability of development of any single crystal,
$p$, where
$$
p = 1-exp (q x \nu), \eqno(1)
$$
\noindent
where $q$ is a parameter that depends only on development conditions,
$x$ is the path length of the ionizing particle through the crystal, and
$\nu$ is the number of ionization acts per unit length.  One could then
write an expression relating the average probability of development
$\bar p$ taken over all crystals in the path under consideration, to the
value of the ionization.  By definition, ${\bar p} = n_g/n_t$, where
$n_g$ (the grain density) and $n_t$ are the number of {\bf developed}
AgBr crystals per unit path length and the {\bf total} number of AgBr
crystals per unit path length, respectively.  If all the grains are
assumed to be of equal diameter $\bar x$ with their centers aligned,
then the average value of the probability is
$$
{\bar p} = 1 - exp (-y), \eqno (2)
$$
\noindent
where $y = q{\bar x} \nu$.  If one then assumes with Demers$^{44}$, that
the AgBr crystals are of equal diameter ${3\over 2}{\bar x}$,
distributed at random about the particle, then
$$
{\bar p} = 1 - {8\over 9}y^2 \left[1 - ({3y\over 2} + 1) exp(-{3y\over
2})\right]. \eqno (3)
$$
\noindent
A further refinement of the crystal distribution consists of replacing
the assumption of equal grain size by the supposition that all values of
crystal diameter between 0 and $2{\bar x}$ are equally probable.$^{45}$
In this case, the mean probability becomes
$$
{\bar p} = 1 - {3\over 2} y^2 \big[(1 - {1\over y}) + (
1 + {1\over y}) exp (-2y)\big]. \eqno (4)
$$
\noindent
We plot the three curves in Fig. 1, and see that they are nearly linear
for small values of ${\bar p}$, but differ as ${\bar p} \rightarrow 1$;
which of the three curves best represents the ionization accurately is
not known as yet.
\vskip 12pt
We then go on to discuss the blob density, $B$, or the number of AgBr
single grains or clusters per unit length of track, and introduce $B^* =
B/B_0$, where $B_0$ is the blob density for minimum ionizing tracks,
related to $n_g^* = n_g/n_0$ by applying appropriate corrections
$^{46}$.  Some authors $^{47}$ recommend the use of total gap length,
$L_H$, while others recommend the use of mean gap length, $\lambda$$^{47,48,49}$.
\vskip 12pt
We go on to describe the experimental apparatus, which was inspired by
the earlier device of Blau, Rudin, and Lindenbaum$^{B62}$ and first reported
by S.C. Bloch$^{50}$. Sylvan Bloch was a graduate student at the
University of Miami who designed the apparatus to be used with a
photomultiplier or phototransistor.  His recollections of Marietta Blau
are given in Appendix III of this paper$^{51}$.  The remainder of the
paper is devoted to comparisons of experimental results of $1.3 GeV/c$
pions and 680 MeV/c antiprotons, and their reaction products, giving
good agreement with the blob density parameter.
\vskip 12pt

In July, 1957, we made an exposure of photographic plates to the 620
MeV/c K$^+-$ meson beam at the Berkeley Bevatron.  This exposure was
unsuccessful because of the fogging of the emulsions from unknown
radiation.  An exposure of a stack of emulsions to the 670 MeV/c
antiproton beam of the Berkeley Bevatron was made in January 1958.  The
results of this study were reported in TN60-461.  Unfortunately, I do not
have a copy of this report, and have not been able to obtain one.  It
was not published in the open literature.
\vskip 12pt
Finally, an exposure to the 450 MeV/c K$^-$-meson beam at the Bevatron
was made in January, 1959.  The interactions of the K$^-$-mesons,
brought to rest in the emulsion, were observed and analyzed.  This work
was presented in a report submitted to the AFOSR$^{B73}$ and later a
part of this work was published in Nuovo Cimento$^{B75}$.  We had
accumulated a substantial number of stopping K$^-$ in emulsion, but
because of the huge statistics accumulated by the European K$^-$
Collaboration,$^{52}$ we restricted ourselves only to describing some
unusual interactions of hyperons.
\vskip 12pt
One event was the likely decay of a hyperfragment in an unusual mode, i.e.,
via a $\pi^{+}$-meson, rather than the customary $\pi^-$-meson.  The
most likely interpretation is a $_{\Lambda} H^{3,4} \rightarrow \pi^+ +
(3,4)n$.  There had been two previous $\pi^{+}$-decay events observed
by Schneps et al $^{53}$.  We quote Deloff et al $^{54}$ and Ferrari et
al $^{55}$, who estimate that on the basis of $\Lambda^0
\rightarrow \Sigma^+$ exchange inside the hyperfragment, the branching
ratio into the $\pi^+$ mode is of order 1\% of the $\pi^-$ mode in the
case of $_{\Lambda}H^{3,4}$.  We discuss also other possible mechanisms for
this phenomenon.
\vskip 12pt
Another unusual event is the possible production and decay of a
$(\Sigma^+ p)$ hyperfragment.  The fast $\pi^+$ meson (about 90MeV)
emitted from the apparent hyperfragment indicates that the short
connecting track is probably a $(\Sigma^+ p) \rightarrow \pi^+ + p + n$,
leading to a Q value not incompatible with that expected for such a
decay, and that the binding energy is less than 1 MeV.  Previous
observations$^{56}$ of such a decay give no more conclusive results than
this one.
\vskip 12pt
We also found several examples of stopping K$^-$ interactions with two
nucleons, which are generally difficult to detect.  We found one example
of K$^- + (pn) \rightarrow \Sigma^- + n + \pi^+$ and another example of
K$^- + (nn) \rightarrow \Sigma^- + \pi^- + p$.
\vskip 12pt
There was likely one, and three other possible examples, of the exotic
decay $\Sigma^+ \rightarrow p + \gamma$, compatible  with a branching
ratio of $1.4 \times 10^{-4}$ to $2 \times 10^{-2}$ to the normal decay
mode, $\Sigma^+ \rightarrow p + \pi^0$, but by no means certain.
\vskip 12pt
There are several apparent events in which the $\Sigma^-$ produced by
the stopping K$^-$ seems to release a large amount of visible energy,
compatible with the decay of a hyperfragment through the interaction
$\Sigma^- + p \rightarrow \Lambda^0 + n$.
\vskip 12pt
There is an unusual apparent decay of a $\Sigma^-$, but the kinematics
strongly suggests the parent particle, with the decay scheme
$(\Sigma^- n)\rightarrow \pi^- + n + n$, again an exotic hyperfragment.$^{57}$
\vskip 12pt
Finally, a sample of four scatterings by $\Sigma^-$ or $\Sigma^+$
interactions in emulsion are added to those found by other groups,
giving a total of ten events.  The total mean free path in emulsion is
then about 38 cm in the energy range from 0 to 200 MeV.  The only
systematic study of the scattering of $\Sigma$-hyperons in bubble
chambers was that of Stannard$^{58}$, who found cross-sections of 38 mb
and 10 mb for the elastic scattering of $\Sigma^+$ and $\Sigma^-$,
respectively, on protons.
\vskip 12pt
Clearly, our work was a significant contribution to the early knowledge
of hyperon interactions.
\vskip 12pt
\leftline{
G. Other Writings by Marietta Blau}
\vskip 12pt
During her years in the United States, Marietta Blau published four more
works, in addition to the research articles I have described above.
\vskip 12pt
The first of these, ``Bericht \"uber die Entdeckung der durch kosmische
Strahlung `Sterne' in photographschen Emulsionen'', was written while
she was at Columbia University$^{B58}$, and published in Sitzungsber.
{\bf 159}, 53-57, (1950).  Blau gives a brief history of her work with
photographic emulsions in the 1920's, her collaboration with
H.Wambacher, and through the friendly support of V.R. Hess (the Nobel
Laureate) and R.
Steinmaurer, exposure of plates at an altitude of 2300 m in 1936.  She
describes the cosmic ray stars found in the emulsions, some with as many
as twelve heavy prongs.  They found that in about 10\% of the cases, the
primary energy was greater than 200MeV.  They were encouraged by these
results, and obtained a grant from the Academy of Sciences through the
efforts of Stefan Meyer for new high altitude exposure.  In a classic
bit of understatement, she says that all this work was interrupted by
the political conditions in Austria.  She concludes the article with a
remark that seems somehow tinged with envy:  ``Meanwhile, this work was
carried out in England and America with improved methods, and the
progress of recent years in theoretical and experimental areas have not
only nearly completely solved the problem of ``Stars'', but have led
also to important knowledge in the fields of cosmic rays and nuclear
physics.''
\vskip 12pt
In the same year, Blau published an article in an apparent festschrift
in honor of Karl Przibram's 70th birthday, entitled, ``M\"oglichkeiten
und Grenzen der photographischen Methode in Kernphysik und Kosmischer
Strahlung,'' in Acta Phys. Austriaca {\bf 3}, 384-395 (1950)$^{B61}$.
\vskip 12pt
This paper, like the previous one but longer, is also a review of the
evolution of the photographic emulsion method in particle and nuclear
physics.  She discusses first the observation of protons in emulsion and
then the observation of neutrons through the recording of their knock-on
protons up to 13MeV.  She describes the work of M. Goldhaber, who
impregnated emulsions with various elements and observed nuclear
reactions.  Emulsions of lower sensitivity, impregnated with uranium
salts or the salts of other radioactive elements, are well suited for the
study of decay processes, and certain important contributions were made
with the emulsion technique.  In 1947, Demers and Wottan observed alpha
particle decay, and in the same year, using the methods of Green and
Livesey and of Tsien-Sam-Tsiang, found threefold and fourfold splitting
of uranium atoms from neutron bombardment.  She goes on to credit first
Ilford Ltd., London and then a short time later Eastman-Kodak, Rochester
with providing a larger content of AgBr and relatively small grains, in
pellicle thickness up to 600 microns.  The development methods were
worked out by Occhialini, Powell and Lattes, who also standardized the
photographic material.  (She singles out also the efforts of P.Demers,
N.A. Perfilov, M.M. Shapiro, H. Yagoda and A. Zhdanov among the other
important contributors to the development of this process).
\vskip 12pt
Blau goes on to discuss the importance of new methods and refined
measuring devices in leading to new discoveries in physics
generally, but especially in the area of nuclear physics.  In the
meanwhile, the predicted particle -- meson -- with about 200 times the
electron mass was confirmed with the Wilson cloud chamber and
Geiger-M\"uller counter.  Corresponding to the previously mentioned
sensitivity of the new emulsions, it appeared possible to obtain meson
tracks up to about 1000 microns to their stopping point.
\vskip 12pt
>From this point on, it is only possible to describe the advances in the
photographic method in connection with the theory of mesons.  Both
developed with each other; the new theoretical knowledge called for
newly refined measuring techniques and, furthermore, emulsions of
greater sensitivity.
\vskip 12pt
In January 1947, Perkins exposed Ilford (B1)-emulsions for some hours in
an aircraft at an altitude of about 10km.  He found in his plates a new,
until then unfamiliar phenomenon -- multiple nuclear disruption by mesons:
 in one star, consisting of four prongs, one of the tracks corresponded
to the slowing down incoming particle with increasing grain density.  It
was determined by measurements that the particle had a mass
substantially less than that of the proton.  Further searches by Perkins
and the physicists of the Bristol group in the cosmic radiation
confirmed the first example, slowing down charged particles of small
mass, present in the cosmic radiation, can disrupt atomic nuclei with
the emission of heavier particles.
\vskip 12pt
Are these light particles identical to the mesons found in the Wilson
chamber?  To decide this question, one must make grain density and
scattering measurements, leading to a mass of about 300 electron masses.
 She compliments the group of Perkins, Lattes, Occhialini and Powell for
pursuing, and finding in the cosmic rays the spontaneous decay of a
$\pi$-meson into a secondary $\mu$-meson.  From the fact that the
$\mu$-mesons had equal lengths, one could determine the masses of the
$\pi$- and $\mu$-meson, the latter accompanied by a neutrino.  The ratio
of masses of the $\pi$- and $\mu$-meson was found to be ${m\pi\over
m_{\mu}} = 1.32$, in agreement with the values found in California from
magnetic measurements on artificial mesons produced by their cyclotron.
\vskip 12pt
She discusses the further increase of sensitivity of emulsions to fast
$\beta$-particles obtained by Eastman-Kodak at the end of 1948 and their
exposure of the Bristol group in the Jung-fraujoch.  They found
interesting cases of $\mu$-meson to $\beta$-particle decays and showers of
$\beta$-particles accompanied by heavy particles.
\vskip 12pt
But besides these discoveries, other researchers in cloud chambers
demonstrated the existence of a charged particle with a mass equal to
half that of a proton.  In the new emulsions the existence of this
meson--$\tau$-meson--with a mass of about 1000 electron masses was
established, and at the end of its track produces a star consisting of
three prongs.  One track is a $\pi$-meson, while although the other two
do not end in the emulsion, they are also identified as $\pi$-mesons,
and, furthermore, all three tracks were coplanar.  Another fortunate case
was observed by Leprince-Ringuet, who observed a stopping $\tau$-meson
that showed it produced in its decay, a mass of more than 700 electron
masses.
\vskip 12pt
Before closing the chapter on ``Pioneering Work of the Photographic
Method'', she devotes several pages to the research of Bradt and Peters
in Rochester, and Frier, Lofgren, Ney, and Oppenheimer in Minnesota, who
exposed plates in balloons at 30 km and found extraordinarily thick and
long tracks.  Using new techniques for analyzing these tracks, they found
them to be the nuclei of various atoms:  iron, sodium, magnesium,
silicon, potassium, calcium.  The energy of observed particles is
extraordinarily large, often greater than a GeV.
\vskip 12pt
The paper concludes with a discussion of grain density and the
range-energy relation.  This portion is clearly based on her paper on
``Grain Density in Photographic Tracks of Heavy Particles''$^{B54}$,
which was done at about the time she was preparing the present article.
\vskip 12pt
The third paper in this group was written by Marietta Blau in 1959 while
she was at the University of Miami, and is titled ``Ionizationmessungen
in photographischen Emulsionen'' and published in Acta Phys.
Austriaca$^{B70}$.  Like the previous paper in this section, it was for
Karl Przibram's festschrift, but this time for his 80th birthday.  This
paper is based in part on our paper discussed above on ``Studies of
Ionization Parameters in Nuclear Emulsions''$^{B72}$ and probably on
another series of articles she was writing at that time for the volume
edited by Luke C.L. Yuan and Chen-Shiung Wu, Methods of Experimental
Physics, Vols. 5A and 5B, Nuclear Physics, $^{B74}$, which I will
discuss below in the last part of this section.  Hence, I will not
elaborate further on this publication, which is presumably covered in
references (B72) and (B74).
\vskip 12pt
The contributions of Marietta Blau to the volumes of Yuan and Wu on
Methods of Experimental Physics, Nuclear Physics$^{B74}$, Vols. (5A and 5B)
are a masterful discussion of the photographic emulsion technique in
nuclear and particle physics by the person who is most responsible for
its discovery and for much of its evolution and application.
\vskip 12pt
The contributions are in five parts.  The first in Vol. 5A, Section
1.7, is the longest, and starts with a historical introduction (1.7.1),
some of which was discussed earlier, and which I will not repeat.  As in
reference (B61), she attributes the discovery of the $\pi^-$-meson to
Perkins$^{59}$ and the positive counterpart to Lattes et al$^{60}$.  The
first heavy meson was the $\tau$-meson discovered by Brown et al$^{61}$
in nuclear emulsions.  She notes that the method also has been
successfully applied in the field of slow neutrons,
photo-disintegrations, and in problems connected with fission.
\vskip 12pt
In the following sections she discusses questions that are clearly
addressed to the concerns of emulsion practitioners, namely (1.7.2) on
Sensitivity of Nuclear Emulsions, (1.7.3.1) on Processing Techniques of
Nuclear Emulsions, with clear recipes for research workers, (1.7.3.2) on Water
Content of Emulsions, and (1.7.4) on Optical Equipment and Microscopes.
\vskip 12pt
In (1.7.5) Blau discusses in some detail the Range of Particles in
Nuclear Emulsions, beginning with (1.7.5.1), Measurement of the Residual
Range of Particles in Nuclear Emulsions, and then (1.7.5.2), Range-Energy
Relations in Nuclear Emulsions, where she introduces the Bethe-Bloch
equation, the fundamental relation describing the energy loss of
particles as function of velocity$^{62}$.  In (1.7.5.3) she discusses
Range Straggling and in (1.7.5.4) the Range-Energy Relations for
Multiply Charged Particles, quoting the seminal work by Barkas$^{63}$.
In (1.7.6) she discusses Ionization Measurement in Emulsions, basing it
on Sternheimer's$^{64}$ adaptation of the Bethe-Bloch equation to
photographic emulsions.  This section and the following one (1.7.7),
Ionization Parameters:  Blob Density, Gap Density, Mean Gap Length and
Total Gap Length, include many of the results of our 1960 paper on
Ionization Parameters$^{B72}$.
\vskip 12pt
In section (2.1.1.3) Blau discusses the Charge Determination of
Particles in Photographic Emulsions.  There are two magnetic methods,
namely the ``sandwich method'' and the magnetic deflection in the
emulsion itself.  In the ``sandwich method'', the measurement is made of
deflections of particles traversing the air gap between two parallel
emulsion sheets.  The curvature of the trajectory of the particle in the
magnetic field existing in this gap is determined by the angle between
the particle's exit and entrance directions in the two adjacent
emulsions.  Although the accuracy of this method could be increased by
using  wider gaps, the maximum separation is limited by the fact that it
becomes increasingly difficult to follow a track.  Distortion may also
impede the usefulness by causing large errors, especially for tracks
with dip angles exceeding 10$^{\circ}$.  However, the method can be used
to obtain at least the sign of the charge.
\vskip 12pt
The other method of magnetic deflection requires distortion free
emulsions.  It yields an accurate value of the charge only if the
magnetic field is large, the path in the field long, and the mean
scattering angle ${\bar \alpha}_{sc}$ is small compared to the magnetic
deviation angle $\alpha_m$, where $\alpha_m = (t/\rho)$, where $\rho$ is the
radius of curvature of a trajectory which describes a path of circular
arc $t$ in a region of field $H$.  Then $\alpha_m = ({tHz\over p})$, where
$z$ is the charge of the particle and $p$ is its momentum.  The authors
in this field have adopted, as a general rule, that the sign of charge
of particles in energy range 200-2000MeV can be determined with 80\%
probability if the path length is 2 cm and the applied magnetic field is
30,000 gauss. I tried myself to apply this method in 1960 with Derek
Prowse at U.C.L.A., but we were met by disappointing results.
\vskip 12pt
>From the previous discussions on ionization, it was seen that the grain
density was expressed by the relation $g = f(z^2\beta)$.  Thus the grain
density alone gives no information on $z$ or $\beta$ separately.  One
must then supplement such observations with an independent method, such
as scattering, which is a measure of momentum and charge.
\vskip 12pt
Another important parameter for the magnitude of charge determination is
the production of $\delta$ rays.  These arise from collisions with
atomic electrons in the emulsion when the energy transfer is greater
than the average energy given to grains forming the particle trajectory.
Thus, the atomic electrons acquire considerable velocities, and hence are
able to render several grains developable,  thereby forming short
trajectories which protrude from the original trace.  $\delta$-ray
measurements were first used to discriminate among charges of energetic
heavy primaries which were discovered in the cosmic radiation by Bradt
and Peters, and by Freier et al$^{26}$.  Blau then gives an extensive
discussion of the application of the $\delta$-ray technique.
\vskip 12pt
In Section (2.2.1.1.5) Blau discusses Momentum Measurement in Nuclear
Emulsions.   This portion of the paper, about twenty pages long, is about
as definitive a treatment as I have seen in the literature on the
subject of multiple scattering of charged particles through matter,
giving an estimate of ($p\beta$), where $p$ is the momentum and $\beta$
is the velocity.  She quotes all the important references in the
literature, and also goes into experimental details.
\vskip 12pt
In (2.2.3.8) Blau discusses the Detection and Measurement of Gamma Rays
in Photographic Emulsions.  She points out that gamma radiation
generally causes a general blackening of the emulsions, similar to that
of visible light, and is applied qualitatively to biological problems.
In physics it becomes more interesting for high energy radiation when
electron pair production sets in and electromagnetic cascades start to
develop.  The study of these cascades, their multiplication, and the
energy of individual electron pairs are used for the determination, or
at least, estimate of the primary photon energy causing these phenomena.
 She discusses the opening angle of electron pairs, their energy
measurement, the ionization method, and finally the problem of
$\pi^0$-meson production and its short decay time, about the limit
that can be determined by visual measurements.  It was an early great
achievement of the emulsion technique.
\vskip 12pt
Finally, in (2.3.5), in volume 5B of the Yuan and Wu, Methods of
Experimental Physics, Blau discusses Determination of the Mass of
Nucleons in Emulsions.  She describes, in some detail, the application
of the constant sagitta method, then the Lund photometric method, the
so-called mean track width method, and then the mass measurement of
particles which do not end in the emulsion.
\vskip 12pt
Had Blau been in better health, it is clear that her last contribution
in Yuan and Wu's book could have become a classic exposition of the
emulsion technique in high energy physics as a book of its own.  As it
stands, even with the passage of about forty years, it remains an
important contribution to the literature of experimental particle
physics.
\vskip 12pt
A Russian translation of Blau's contributions was apparently issued in 1965.$^{B75}$
\vskip 24pt
\leftline{H. Epilogue}
\vskip 12pt
In April 1960, Marietta left by train from Miami, to New York, and then
on to Vienna.  My family and the Carters bade her an emotional farewell.
\vskip 12pt
I continued to work on our K$^-$-interactions results, and then obtained
a grant from the Research Corporation to expose a stack of Ilford K5
emulsion plates to the 800 MeV/c K$^-$-meson beam at the Bevatron in
Berkeley.  I left Miami in the Summer of 1961 for the University of
Trieste, where I had hoped to collaborate with Carlo Franzinetti.
Unfortunately, he had just left for the University of Pisa, so I was
left to work on the emulsions by myself, and to collaborate with the
bubble chamber group at Trieste which was scanning photographs of the
antiproton beam at CERN.  Since the scanning and measuring devices for
the bubble chamber photographs were still in development, I collaborated
with a young colleague at Trieste, Mario Ceschia, to perform
phenomenological computations on antiproton-proton interactions at
laboratory energies from 30 to 180 MeV.  We based our analysis on a
refinement of the Ball and Chew model$^{65}$ and on the adaptation of
modifications of the $p-p$ potential to the ${\bar p} - p$ potential, and
found decent agreement with the experimental results then
available,$^{66}$ also making predictions on differential cross sections and
polarization.
\vskip 12pt
With the help of three scanners, I also found two heavy hyperfragment
decays produced by the fast K$^-$-beam$^{67}$.  My stay in Trieste was
supported by I.N.F.N. (Istituto Nazionale Fisica Nucleare).
\vskip 12pt
In the Spring of 1962, my family and I traveled from Trieste to Vienna
where we met with Marietta Blau at her apartment.  She was, of course, a
gracious hostess, but the main purpose was to finish our hyperon
paper$^{B75}$.  She expressed great displeasure at her treatment by the
faculty at the Institut for Radiumforschung, and her inability to
engage in productive work.  She was clearly not in the best of physical
health.
\vskip 12pt
I extended my leave from the University of Miami, by going to the
Weizmann Institute of Science in Rehovoth, Israel, in August 1962, with
a grant from the Israel Atomic Energy Commission.  There I collaborated
with a group that had been studying the interactions of K$^-$- mesons
with emulsion nuclei with the production of ($\Sigma^{\pm} +
\pi^{\mp}$).  We combined their results with ours from Miami and carried
out a detailed analysis of the results.  We found that the coulomb
effect on the energies of the charged $\Sigma$-hyperons and charged
$\pi$-mesons was observable, and had an effect on the $\Lambda(1405)$
resonance$^{68}$.
\vskip 12pt
In 1963 I returned to the University of Miami, where I collaborated with
Joseph Cox on a phenomenological treatment of particle
scattering$^{69}$.  In the late 1960's, I collaborated with a group from
Northwestern University on experiments using spark chambers designed to
measure the magnetic moment of the $\Sigma^+$-hyperon at Argonne
National Laboratory.  In the 1970's and early 1980's I collaborated with
a group at the University of Michigan which carried out experiments on
high energy polarized proton-proton scattering at Argonne National
Laboratory and Brookhaven National Laboratory,  where the particle
detectors were scintillation counters.
\vskip 12pt
I mention these endeavors mainly to emphasize my debt to the influence
and legacy of Marietta Blau over the rest of my life.  When I first read
the article by Peter Galison in Physics Today in 1997$^4$, I was
prompted to send a short letter to Physics Today in 1998$^{70}$.  In it
I wrote that my several years of working with Marietta had changed my
life.  After rereading the present article, I can emphatically reaffirm that
remark and express gratitude for my association with that wonderful
person.  She led me into particle physics, and even if I am dissatisfied
with the quality of my achievements, I am grateful for being a
participant in interesting experiments, and in having established a
world-wide circle of friends and associates in this exciting field.
\vskip 12pt
Actually, these associations were greatly amplified after 1964, when my
colleague at the University of Miami, the theorist, Behram Kursunoglu,
with my help, established the Center for Theoretical Studies and the now-famous
Coral Gables Conferences on Symmetry Principles at High Energy, and which
have persisted, with some name changes, to this day.  The Center was host
to some of the great scientists of our time, including P.A.M. Dirac, R.
Feynman,
Lars Onsager, J.R. Oppenheimer, E. Wigner, H. Bethe, F. Crick, E.
Teller, R. Hofstadter, M. Gell Mann,  W. Lamb, J. Schwinger, A. Salam,
S. Weinberg, V. Zworykin, among others, and also to a number
of gifted post doctoral fellows.  After Kursunoglu's retirement in 1992,
we have continued these activities under the auspices of the Global
Foundation, and I continue to collaborate with him on problems of
unified field theory and astrophysics.
\vskip 12pt
I return now to the subject that must have been the source of great pain
and frustration in Marietta Blau's professional life, namely the
official neglect of her role in the discovery of the pion that I
mentioned at the beginning of this article.  She was too proud and
private to reveal this disappointment to me openly, but I do recall that
she had great disdain for C.F. Powell.  This is corroborated by reading
her Acta Phys.Austriaca article$^{B61}$ and her article in Yuan and
Wu's book$^{B74}$, where she actually attributed the discovery of the
$\pi^-$-meson to Perkins$^{59}$ and that of the $\pi^+$-meson to Lattes,
Occhialini and Powell$^{60}$.  I now remember that she thought more
highly of Perkins, Lattes and Occhialini than she did of Powell, and
felt that the latter was rewarded more for his public posture than for
his achievements or abilities.
\vskip 12pt
I think that it is clear from some of the work described in this paper
and the summary in Appendix IV, that the pion was the central
discovery in nuclear physics after the war.  It is interesting to
recount briefly the history leading up to this discovery.
\vskip 12pt
In 1935, Hideki Yukawa, the great Japanese theorist, attributed the
finite range ($\sim 10^{-15}m$) of nuclear forces between nucleons to the
existence of a massive field quantum or carrier of integral spin$^{71}$.
This has turned out
to be a fundamental concept in the theory of fields, and has led to
further application in weak interactions.  The estimate of the mass of
this carrier was about 300 electron masses.  There ensued then one of the
great nasty tricks that nature plays on the mortal explorers of its
secrets.  In 1937, C.D. Anderson$^{72}$ observed what was called the
$\mu$-meson, or muon, in cloud chambers exposed to the cosmic rays.
Anderson, incidentally, had first also observed the positron, or the
antielectron, several years earlier, after it had been postulated by
P.A.M. Dirac$^{73}$.  The muon was found to have a mass of about 206
electron masses, and furthermore it did not interact strongly in nuclear
matter as should the Yukawa particle.  Hence, the nuclear physicists
were faced with the conundrum.  The muon was actually also a lepton, or,
a ``heavy electron''.  Thus it was that in 1947,with the discovery of
the pion by Perkins, Lattes, et al that the dilemma was resolved.  It
was suggested first by Marshak and Bethe$^{74}$ that two particles were
involved, and indeed the experiments showed that $\pi^+\rightarrow \mu^+
+\nu_{\mu}$ and $\pi^-\rightarrow \mu^- + {\bar \nu}_{\mu}$, and that the
pion has a mass of about 274 electron masses, as I have mentioned in
Appendix IV.  Thus the pion was the ``prize plum'' that ushered in the
great era of high energy physics of the last half of the 20th century.
\vskip 12pt

Marietta Blau's contributions were central to this discovery, but as I
said in the early part of this paper, after the end of the war she was
no longer in a position to forge new roads, but still made contributions
at a modest level.  It is my fervent belief, that had Blau been able to
expose her photographic plates in 1938, in the high Alps, with Stefan
Meyer's sponsorship, that she would have been able to detect a
$\pi$-meson.  This is a poignant example of how social and political
upheaval  can thwart the purest quest for new knowledge.
\vskip 12pt
As an aside, it is truly remarkable how quickly new devices and
techniques are developed.  As I mentioned earlier, emulsions were
largely supplanted by bubble chambers in the 1960's, those in turn by spark
chambers, then streamer chambers, and now whole new classes of
detectors.  It is a further irony that at present, high energy physics
is again turning toward outer space, with the advent of new detectors in
space and on earth.  I feel privileged to be a witness to these events.
\vfill\eject
\centerline{Acknowledgements}
\vskip 12pt
I am grateful to my wife, Lynn Meyer, for her encouragement and for her
skillful criticism of this manuscript.
\vskip 12pt
I would also like to thank Judy Mallery for her help in typing this
manuscript.
\vfill\eject
\centerline{Appendix I}
\centerline{Recollections of Marietta Blau}
\centerline{by Martin M. Block, Northwestern University}
\centerline{Evanston, IL  60208}
\vskip 24pt
Indeed, I did work with Marietta Blau at Columbia, in the late 40's.  I
was a graduate student and she was a Research Associate.  As a graduate
student, I was in charge of the nuclear emulsions section.  Of course,
none of us knew anything about emulsions---it took Blau to teach us the
techniques of developing, scanning, etc.  In this, she was
indispensable...We were still building the Nevis Cyclotron (in those
days, grad students really built the equipment, tested it, etc., since we
had no engineers or post-docs), and for practice, exposed some nuclear
emulsions in balloons.  We came across a strange event, which we
published in '49.  It took until 1999 for me to work again in cosmic
rays---I just published a paper in PRL on Dec. 13 (TODAY).  I can't
tell you anything about her Brookhaven work.  I do know that she was
treated rather shabbily at Columbia---but who wasn't during that era.
\vfill\eject
\centerline{Appendix II}
\centerline{My Interactions with Marietta Blau}
\centerline{by S.J. Lindenbaum, Brookhaven National Laboratory}
\centerline{Upton, NY.}
\vskip 24pt
After the war I simultaneously was employed as a full time staff member
of the ``Nevis Cyclotron Laboratories'' of Columbia University
(1946-1951) and was pursuing my Ph.D. in Physics at Columbia.
\vskip 12pt
I was initially assisting Professor Rainwater (in effect a Deputy
Director) under Professor Booth (the Director) in developing the Radio
Frequency System for the Nevis approximately 400 MeV Cyclotron design
and construction project.  However, my duties were soon broadened in this
small scientific staff to include most aspects of the Cyclotron Project
including planning elements of the future research program.
\vskip 12pt
Marietta Blau was a senior staff member brought in to prepare an
emulsion program for the Cyclotron, as she was the well known
outstanding pioneer in the emulsion technique and its application to
studying Nuclear Stars.
\vskip 12pt
I was asked to look into her program and help where I could.  I found
Marietta to be almost unbelievably well versed in every aspect of the
emulsion techniques even though my expectations were high, considering
her reputation as the outstanding pioneer that she was.
\vskip 12pt
She set up a very complete emulsion laboratory, with scanners that she
trained herself.  Marietta also taught me the emulsion technique at a
sophisticated level which was unattainable through reading the
literature, which I did.
\vskip 12pt
It was obvious to her that the high data rates an FM Cyclotron would
generate in a long program would overwhelm human scanners, and she felt
that as much automation as possible should be developed for the scanning
and analysis programs, and this would also eliminate most of the biases
introduced by human scanning.  I totally shared her conclusion, and
since I guessed that the first Nevis Cyclotron experiment would be most
rapidly done with emulsions, and I had a thesis topic in mind-- namely to
solve a great uncertainty of the times-- what was the process or
processes which lead to Cosmic Ray stars.
\vskip 12pt
Thus I began to collaborate extensively with Marietta and an engineer R.
Rudin, on the development of a semi-automatic device for analyzing
events in nuclear emulsions.  This program was successfully completed and
published [Rudin, R., Blau, M., Lindenbaum, S.J., Semi-Automatic Device
for Analyzing Events in Nuclear Emulsions. Rev. Sci Instr.21,978-985
(1950), and Phys Rev. 78, 319A (1950)].
\vskip 12pt
I believe if emphasis had not quickly shifted to the bubble chamber,
and soon thereafter to the electronic ``bubble chambers'', which I later
pioneered, that this semi-automatic device would have become quite
important.  In any event, it was the first idea which was successfully
developed to automate visually observed events, and likely influenced
future developments.  Marietta Blau, in addition to being an outstanding
scientist, was a warm and caring individual whose company was always
very much enjoyed by myself, and her colleagues, and although a quiet
individual, she certainly stimulated me and other colleagues greatly.
\vskip 12pt
I went on to do my thesis--the first Nevis experiment - by exposing
minimum ionization track sensitive emulsions to internal 350-400 MeV
proton beam and demonstrated, that contrary to the most popular belief
at the time--that the Fermi Statistical Theory explained the major
mechanism of Nuclear Star Formation -- that actually a cascade of
individual nucleons was the major element in Star production, and that
the only role of the Fermi Statistical theory was to cause a
thermodynamic evaporation due to rearrangement of the holes in the Fermi
sea produced by the nucleonic cascade [Bernardini G. Booth E.T.
Lindenbaum S.J., Phys. Rev. {\bf 80}, 905 (1950), Phys. Rev. {\bf 83}, 669-671, (1951),
Phys. Rev. {\bf 82}, 307 (1951), Phys. Rev. {\bf 85}, 826-834, (1952),
Phys. Rev. {\bf 88},
1017-1026 (1952).
\vskip 12pt
Although Marietta Blau did not participate in this extensive important
program, it was the laboratory she developed which was used for it.
\vfill\eject
\centerline{Appendix III}
\centerline{``Recollections of Marietta Blau at Miami}
\centerline{ by S.C. Bloch, Professor Emeritus}
\centerline{ University of South Florida, Tampa, Florida}
\vskip 24pt
I was privileged to be one of Professor Blau's graduate students --
perhaps her last one.  Under her direction, I earned a Master's degree
while participating in her research project at the University of Miami,
funded by the Air Force Office of Scientific Research.  The research was
based on the use of nuclear emulsions to study fundamental particle
interactions.  This work resulted in one of my first publications, which
was one of Professor Blau's last publications:
\vskip 12pt
\itemitem{} ``Studies of Ionization Parameters in Nuclear Emulsions'' in
Review of Scientific Instruments {\bf 31}, 289-297 (1960), M. Blau, S.C.
Bloch, C.F. Carter, and A. Perlmutter.
\vskip 12pt
I was also a student in Professor Blau's course in nuclear physics.  She
was an excellent teacher, very demanding, and respected by her students.
As in her research, she had high standards and expected the best from
students.
\vskip 12pt
Professor Blau was no less demanding from her colleagues.  I was in her
office one day when she was berating another professor, saying, ``You know
nothing about fundamental particles!''
\vfill\eject
\centerline{Appendix IV}
\vskip 12pt
A Brief Tutorial on Units and Nomenclature in Particle Physics.
\vskip 12pt
The masses and charges of elementary particles have been established by
a variety of methods over the course of the past century.  For example,
the mass of the electron is about $m_e = 9.11 \times 10^{-31}kg$  and its
charge is $q_e = -e$, where $e = 1.60\times 10^{-19}$ coulomb.  The mass
of the proton is $m_p = 1.67 \times 10^{-27}kg$ and its charge is $q_p =
+e$.  As far as we know, these are the only stable elementary particles,
and are, along with the neutron, which has no charge and a mass slightly
larger than that of the proton, the principal constituents of the atom.  All other known particles have charges which are either neutral
($q=0$) or are positive or negative multiples of $e$, with the exception of
quarks, which I will discuss later.
\vskip 12pt
As for masses, it is much more convenient to make use of Einstein's
famous relation, $E_0 = mc^2$, where $E_0$ is called the rest energy,
$m$ is the mass of the particle in $kg$, and $c = 3.00 \times 10^8 m/s$
is the approximate speed of light.  This allows one to introduce another
energy unit, the electron volt (eV), where $1eV = 1.60 \times
10^{-19}$Joules.  Then, the electron has a rest energy of about $E_{oe}
= 0.511 \times 10^6eV = 0.511$MeV, and the proton a rest energy of about
$E_{op} = 938.3$MeV, or about 1840 times that of the former.  The free
neutron has a rest energy of about $E_{on} = 939.6$MeV and decays
(weakly) in about 14.8 minutes to a proton, an electron and an
anti-electron neutrino, i.e., $n\rightarrow p + e^- + {\bar \nu}_e$.
The neutron is stable only when it is bound in an atomic nucleus.
\vskip 12pt
The principal units utilized in giving energies or mass energies, are
$KeV(10^3eV)$, $MeV(10^6eV)$, $GeV(10^9eV)$ and $TeV(10^{12}eV)$.  Prior to
the 1960's $BeV$, was used for $10^9eV$, but since the $B$ stood for
``billion'' in the U.S.A., and ``billion'' in Britain meant $10^{12}$, it
was decided to represent $10^9$ by the prefix Giga, hence GeV.
\vskip 12pt
The total energy of a particle is given by $E = K + E_0$, where $K$ is
the kinetic energy.  The relation $E = \sqrt{p^2c^2 + E_0^2}$, where $p$
is the momentum, is very useful.  Here, it is seen that the unit for $p$
is conveniently $(eV/c)$ where $c$ is the speed of light.
\vskip 12pt
The terminology used to identify various particles, radioactive nuclei
and radiation has undergone some changes over the years.  But the early
designation by E. Rutherford at the beginning of the last century of the
various radiations emanating from radioactive nuclei as $\alpha, \beta,
\gamma$ are still in use.  $\alpha$-radiation refers to the nuclei of
helium atoms, with charge +2e and a mass nearly four times that of a
proton.  Hence, $\alpha$-particle is synonymous with helium nucleus
$^4_2He$.  $\beta$-radiation refers to electrons, of charge (-e) and
mass $m_e$ emitted from nuclei--identical with the electrons in the
outer regions of the atom.  Since the discovery of the electron's
antiparticle, the positron, (e$^+$) whose charge is +e and mass $m_e$,
the term $\beta$ is generic for both e$^-$ and e$^+$.  The third kind of
radiation, $\gamma$, refers to gamma radiation, high energy photons, of
no mass or charge, and with energies from a few KeV up to many TeV.
Actually, $\gamma$ is a generic term for photon, which can also refer to
radio, microwave or visible light.  In nuclear and particle physics,
gamma rays emanate from nuclei, as contrasted with x-rays, which are
produced by transitions of closely bound atomic electrons.  Generally speaking,
gamma rays are more energetic than x-rays.
\vskip 12pt
Here, it might be useful to consider some of the principle elementary
particles.  The $\pi$-meson, or pion, comes in a family of three, $\pi^+,
\pi^-, \pi^0$, where the rest energies are about $m_{\pi^{ \pm}} = 139.6$MeV
and $m_{\pi^0} = 135.0$MeV, and decays according to the schemes,
$\pi^+\rightarrow \mu^+ + \nu_{\mu}$, and $\pi^-\rightarrow \mu^- +
{\bar \nu}_{\mu}$, both in about $2.6 \times 10^{-8}$ seconds, and
$\pi^0\rightarrow \gamma + \gamma$ in about $8.4 \times 10^{-16}$
seconds ($\gamma$ is the symbol for massless photon).  The
$\mu^{\pm}$-lepton (originally called $\mu$-meson, and now
muon) is some times called the ``heavy electron'', has a mass of about
$105.7$MeV, and decays in about $2.2 \times 10^{-6}$ seconds, according
to the schemes $\mu^+ \rightarrow e^+ + \nu_e + {\bar \nu}_{\mu}$ and
$\mu^- \rightarrow e^- + {\bar \nu}_e + \nu_{\mu}$, where $\nu_e$ and
$\nu_{\mu}$ stand for two kinds of neutrinos, both of which were
considered to be massless until the past few years, when underground
experiments at Kamiokande in Japan, and elsewhere, indicate that
neutrinos have a very small rest energy of about or less than 1eV, thus
supplanting the electron as the lightest particle (photons, of course, have
zero rest energy, or at best $<3. \times 10^{-27}eV$).  There is
another lepton, discovered only in the 1970's, $\tau$-lepton, with a
rest energy of about 1777.1 MeV and which decays via many modes, in
about $2.96 \times 10^{-13}$ seconds, all involving a $\nu_{\tau}$ or
${\bar \nu}_{\tau}$, to muons, and to various combinations of
pions and kaons.  It also comes with positive or negative charge $\pm
e$.
\vskip 12pt
The kaon (or K-meson) called a ``strange'' particle, discovered shortly
after the pion in 1948, comes in several varieties, $K^+,K^-,K^0,{\bar
K}^0$, and has rest energies of about $E_{0K^{\pm}} = 493.7$MeV and
$E_{0K^0} = 497.7$MeV.  The lifetime of $K^{\pm}$ is about $1.24 \times
10^{-8}$ seconds and has many decay schemes, including $K^+\rightarrow
\mu^+ + \nu_{\mu},~K^-\rightarrow \mu^- + {\bar \nu}_{\mu}$, and also
$K^{\pm}\rightarrow (3\pi)$ (originally called a $\tau^{\pm}$-meson) and
$K^{\pm}\rightarrow (2\pi)$ (originally called a $\theta$-meson).  The
$K^0$ comes in two principal varieties, $K^0_s (K^0$-short) and $K^0_L
(K^0$-long), with lifetimes of about $0.89\times 10^{-10}$ seconds and
about $5.2 \times 10^{-8}$ seconds.  They oscillate between each other,
and are responsible for the observation of time reversal non-invariance.
\vskip 12pt
The hyperons, which first appeared in the late 1940's, 1950's and
1960's, are also called strange particles and all have rest energies
larger than that of the proton.  The lightest of these is the
$\Lambda^0 (1115$MeV), then a family called $\Sigma$ comprised of
$\Sigma^+(1189$MeV), $\Sigma^-(1197$MeV), $\Sigma^0(1192$MeV).  The
$\Sigma^+$ and $\Sigma^-$ decay mainly into nucleon (n or p) and pion
($\pi^+, \pi^-,$ or $\pi^0$) in a time less than $10^{-10}$ seconds,
while the $\Sigma^0$ decays in about $10^{-19}$ seconds into a
$\Lambda^0$-hyperon and a photon.  The heaviest of these hyperons
is the $\Xi$, which appears in two charge states, the
$\Xi^0 (1315$MeV) and $\Xi^-(1321$MeV), and decay mainly into
$\Lambda^0$ and $\pi^0$ or $\pi^-$ in times of about $10^{-10}$
seconds.  The six hyperons named here, and the two nucleons, proton
$(p)$ and neutron $(n)$ form an octet of particles in one of the early
classification schemes established by M. Gell Mann in 1961.  In the same
spirit, the mesons $\pi^+, \pi^-, \pi^0, \eta, K^-, {\bar K}^0, K^+,
K^0$, form the lowest octet in this scheme, where the $\eta$ is a
neutral meson resonance of rest energy 547 MeV, decaying mainly into
three pions or two photons in about $3.4 \times 10^{-18}$ seconds.
\vskip 12pt
These classification schemes were obtained by introduction of quarks, by
M. Gell Mann and G. Zweig$^{75}$.  They postulated that there are three
quarks $u$(up), $d$(down), and $s$(strange) of charges $-1/3 e, 2/3 e,
$ and $-1/3e$, respectively and their antiparticles, which are the
constituents of mesons (2 quarks) and baryons (3 quarks).  In the 1970's
and 1980's and 1990's there were postulated and confirmed three more
quarks, $c$(charm), $b$(bottom), and $t$(top), which generated an
explosion of information and theoretical progress.  For the purposes of
this paper it is necessary only to mention that the first decuplet of
baryon resonances, which we discussed in section E$^{B69}$, could be
understood as a grouping of $u,d$ and $s$ quarks.
\vskip 12pt
There is a related way of describing particle groupings that preceded
these schemes, introduced by W. Heisenberg$^{76}$ to describe atomic
nuclei, namely isospin (I).  This quantity can take on values I = 0
(singlets, such as $\Lambda^0$, I = 1/2 (doublets, such as $n$
and $p$), I=1 (triplets, such as ($\Sigma^+, \Sigma^0, \Sigma^-$) and I
= 3/2 (quartets, such as ($\Delta^-, \Delta^0, \Delta^+, \Delta^{++}$),
etc.
\vskip 12pt
While quarks have not been seen in a free state, since they are confined
to elementary particles by strong forces involving also gluons, they can
be detected only indirectly, but their utility in bringing an order to
hundreds of particle states has been of inestimable value.
\vskip 12pt
           Originally, mesons were so named because they had masses intermediate
between electron and proton.  Now, with the discovery of heavier mesons
arising from the heavier quarks, c, b and t, mesons may be much heavier
than protons.  More fundamental is the fact that because mesons are
composed of two spin 1/2 quarks, they must have integer spin, i.e., 0,
1, 2, 3... and they are designated as Bose-Einstein particles, or
bosons.  They obey quite different statistics from Fermi-Dirac
particles, or fermions,  which have half-integer spin, i.e. ${1\over 2}$,
${3\over 2}$, ${5\over 2}$,... Baryons, which are composed of three spin
1/2 quarks, are therefore fermions, as are also electrons, muons,
tauons, and neutrinos.
\vskip 12pt
For the reader who wishes to go beyond the early results discussed in
this paper, I can recommend several of many good publications.  One is
the Physical Review, Review of Particle Properties, {\bf D50}, No. 3, (1
August 1994).  This one is the more or less official compendium and
probably contains more information than most people need.
A new update of this should appear shortly.  A more accessible source is
``Introduction to High Energy Physics'', by Donald H. Perkins, 3rd
Edition, Addison-Wesley Publishing Company, Inc. 1987.
\vfill\eject
\centerline{\bf Footnotes}
\item{$^1$} Peter Galison, {\it Image and Logic: A
Material Culture of Microphysics,} University of Chicago Press (Chicago
and London) 1997.
\item{$^2$} Walter Moore, {\it Schr\"odinger, life and thought}
(Cambridge University Press) 1989, pp. 479-480.
\item{$^3$} Max Born, {\it Atomic Physics}, Blackie and Son Ltd.,the
(Glasgow) 1969, pp. 36-37.
\item{$^4$} Peter Galison, {\it Physics Today} {\bf 50}, 42, Nov. 1997.
\item{$^5$} Leopold Halpern in {\it Women in Chemistry and Physics; a
Bibliographical Sourcebook} Edited by L.S. Grinstein, R.K. Rose, and
M.H. Rafailovich, Greenwood Press (Westport, Connecticut and London),
1993, pp. 50-56.
\item{$^6$} Nina Byers, {\it Contributions of 20th Century Women to
Physics}\hfill\break (http://www.physics.ucla.edu/$^{\sim}$cwp/phase
2/Blau\_Marietta@843727247.html).
\item{$^7$} C.F. Powell, P.H. Fowler and D. Perkins, {\it The Study of Elementary
Particles by the Photographic Method}, Pergamon Press (New York,
London, Paris, Los Angeles) 1959.
\item{$^8$} J.B. Birks, {\it The
Theory and Practice of Scintillation Counting}, The MacMillan Company in
(New York) 1964.
\item{$^9$} S.C. Curran and W.R. Baker, U.S. Atomic
Energy Report MDDC, 1296, 17th Nov (1944); Rev. Sci. Instr. {\bf 19},
116 (1948).
\item{$^{10}$} J.W. Coltman and F.H. Marshall,  Nucleonics {\bf 1},
358 (1947); F.H. Marshall and J.W. Coltman, Phys. Rev. {\bf 72}, 528
(1947); F.H. Marshall, J. Coltman and A.I. Bennett, Rev. Sci. Instr. {\bf
19}, 744 (1948); F.H. Marshall, J.W. Coltman and L.P. Hunter, Rev. Sci.
Instr. {\bf 18}, 504 (1947).
\item{$^{11}$} I. Broser and H. Kallmann, Z. Naturforsch. {\bf 29}, 439,
642 (1947); H. Kallmann, Natur und
Technik (July, 1947); Phys. Rev. {\bf 78}, 621 (1950).
\item{$^{12}$} P.R. Bell, Phys. Rev. {\bf 73}, 1405 (1948).
\item{$^{13}$} R. Hofstadter, Phys. Rev. {\bf 74}, 100 (1948).
\item{$^{14}$} H. Petterson, BerIIa, Wien {\bf 132}, 55 (1923).
\item{$^{15}$} R.D. Evans, Nucleonics {\bf 1}, no. 2, 32 (1947).
\item{$^{16}$} C.C. Dilworth, G.P.S. Occhialini, and R.H. Payne,
Nature {\bf 162}, 102 (1948).
\item{$^{17}$} Crabtree, Parker, and Russel, Soc. Mot. Pic. ENO, VO RR
{\bf 21}, 21 (1933).
\item{$^{18}$} J. Rotblat ``Photographic Emulsion Technique'', Prog. in
Nucl. Phys., {\bf 1}, 37-72 (1950).
\item{$^{19}$} C.M.G. Lattes, P.M. Fowler, and P. Cuer, Proc. Phys. Soc.
London {\bf 59}, 883 (1947).
\item{$^{20}$} C.M.G. Lattes, G.P.S. Occhialini, and C.F. Powell, Nature
{\bf 160}, 486 (1947).
\item{$^{21}$} J.H. Webb, Phys. Rev. {\bf 74}, 511 (1948).
\item{$^{22}$} D.H. Perkins, Nature {\bf 159}, 126 (1947).
\item{$^{23}$} P. Debye and E. H\"uckel, Physik. Zeits. {\bf 24}, 185
(1923).
\item{$^{24}$} See Appendix I, ``Recollections of Marietta M. Blau'',
Martin M. Block (Dec. 13, 1999).
\item{$^{25}$} See Appendix II, ``My Interactions with Marietta Blau'',
S.J. Lindenbaum (Feb 21, 2000).
\item{$^{26}$} H.L. Bradt and B. Peters, Phys. Rev. {\bf 74}, 1828
(1948); P. Frier et al, Phys, Rev. {\bf 74}, 413 (1948).
\item{$^{27}$} S.J. Lindenbaum, Columbia University thesis 1951.
\item{$^{28}$} J.B. Harding, Nature {\bf 163}, 440 (1949); Phil. Mag.
{\bf 42}, 63 (1951).
\item{$^{29}$} Menon, Muirhead and Rochat, Phil. Mag. {\bf 41}, 583 (1950).
\item{$^{30}$} P.E. Hodgson, Phil. Mag. {\bf 42}, 955 (1951).
\item{$^{31}$} Camerini, Davies, Franzinetti, Lock, Perkins, and
Yekutieli, Phil.Mag. {\bf 42}, 126 (1951).
\item{$^{32}$} W.F. Fry, Phys. Rev. {\bf 91}, 1576 (1953).
\item{$^{33}$} S.J. Lindenbaum and L.C.L. Yuan, Phys. Rev. {\bf 92}, 1578
(1953).
\item{$^{34}$} M. Sands, Phys. Rev. {\bf 77}, 180 (1950); W.O. Lock and
G. Yekutieli, Phil. Mag. {\bf 43}, 231 (1952).
\item{$^{35}$} A high percentage of stoppings has been  reported for 1.5 GeV
negative pions by W.D. Walker and J. Crussard, Phys. Rev. {\bf 98}, 1416
(1955).
\item{$^{36}$} S.J. Lindenbaum and L.C.L. Yuan, Phys. Rev. {\bf 100},
306 (1955).
\item{$^{37}$} M. Danysz and T. Pniewski, Phil. Mag. {\bf 44}, 348
(1953); W.F. Fry, J. Schneps, and M.S. Swami, Phys. Rev. {\bf 99}, 1561 (1955).
\item{$^{38}$} L.M. Eisberg, W.B. Fowler, R.M. Lea, W.D. Shepherd, R.P.
Shutt, A.M. Thorndike and W.L. Whittemore, Phys. Rev., {\bf 97}, 797
(1955).
\item{$^{39}$} R.R. Crittenden, J.H. Scandrett, W. D. Shephard, W.D.
Walker and J. Ballam, Phys. Rev. Lett., {\bf 2}, 121 (1959).
\item{$^{40}$} E. Fermi, Prog. Theor. Phys. (Japan), {\bf 5}, 570
(1950); Phys. Rev., {\bf 92}, 452 (1953).
\item{$^{41}$} R.M. Sternheimer and S.J. Lindenbaum, Phys.
Rev. {\bf 109}, 1723 (1958).
\item{$^{42}$} This leads to a
nominal mass of the isobar state to be $m_p + m_{\pi} + Q = (940 + 140 +
150)MeV = 1230$ MeV, remarkably close to the mass of the $\Delta$
resonance of 1232 MeV.  Remember that this was several years before the
advent of the quark model and the identification of the lowest lying
decuplet composed of
an $I = 3/2,~ \Delta(1232); I = 1,~ \Sigma(1385); I = {1\over
2},\Xi(1530);~ I = 0,~\Omega^-(1672)$.
The numbers in parentheses are the approximate rest
energies of the $\Delta$ resonances in MeV.
The width of the $\Delta$ resonance is 120 MeV, in agreement with the paper's
result.  As far as I know, this was the first evidence for the $\Delta$
resonance, in this case a $\Delta^{\circ}$, allegedly composed of an up-
and two down- quarks.  Hence, although we were disappointed not to match
the predictions of the isobar model, it is gratifying, in retrospect to
give early evidence for the $\Delta$ resonance.
\item{$^{43}$} N. Metropolis, R. Bivins, M. Storm, J.M. Millerr, G.
Friedlander and A. Turkevich, Phys. Rev., {\bf 110}, 204 (1958).
\item{$^{44}$} P. Demers, Can. J. Research {\bf A25}, 223 (1947).
\item{$^{45}$} P.H. Fowler and D.H. Perkins, Phil. Mag. {\bf 46}, 587 (1955).
\item{$^{46}$} G. Alexander and R.H. W. Johnston, Nuovo Cimento {\bf 5},
363 (1957).
\item{$^{47}$} Castagnoli, Cortini, and Manfredini, Nuovo Cimento {\bf
2}, 301 (1955).
\item{$^{48}$} C. O'Ceallaigh, Nuovo Cimento, Suppl. {\bf 12}, 412 (1954).
\item{$^{49}$} M.G. Menon and C.O'Ceallaigh, Proc. Roy. Soc. (London)
{\bf A221}, 292 (1954).
\item{$^{50}$} S.C. Bloch, Rev. Sci. Instr. {\bf 29}, 789 (1958).
\item{$^{51}$} See Appendix III, ``Recollections of Marietta Blau at
Miami'', S.C. Bloch, (April 3, 2000).
\item{$^{52}$} K$^-$ Collaboration, Nuovo Cimento {\bf 13}, 690 (1959);
{\bf 14}, 315(1959); {\bf 15}, 873 (1960).
\item{$^{53}$} J. Schneps, Phys. Rev. {\bf 112}, 1335(1958); Y.W. Kang,
N. Kwak, J. Schneps and P.A. Smith, Nuovo Cimento {\bf 22}, 1297 (1961).
\item{$^{54}$} A. Deloff, J. Szymanski and J. Wrzeconko, Polish Academy
of Sciences, Institute of Nuclear Research, Report no. 95/VII, (June
1959).
\item{$^{55}$} F. Ferrari and L. Fonda, Nuovo Cimento {\bf 7}, 320
(1958); N. Dallaporta and F. Ferrari, Nuovo Cimento, {\bf 5}, 111 (1957).
\item{$^{56}$} M. Baldo-Ceolin, W.F. Fry, W.D. B. Greening, H. Huzita
and S. Limentani, Nuovo Cimento {\bf 6}, 144(1957); R.C. Kumar and F.R.
Stannard, Nuovo Cimento {\bf 14}, 250 (1959).
\item{$^{57}$} E. Gandolfi, J. Hengheboart and E. Quercigh, Nuovo
Cimento {\bf 13}, 864 (1959).
\item{$^{58}$} F.R. Stannard, Phys. Rev. {\bf 121}, 1513 (1961)..
\item{$^{59}$} D.H. Perkins, Nature {\bf 159}, 126 (1947).
\item{$^{60}$} C.M. G. Lattes, G.P.S. Occhialini and C.F. Powell, Nature
{\bf 160}, 486 (1947).
\item{$^{61}$} R.H. Brown, V. Camerini, P.H. Fowler, H. Muirhead, C.F.
Powell and D.M. Ritson, Nature, {\bf 163}, 82 (1949).
\item{$^{62}$} F. Bloch, Z. Physik {\bf 81}, 363 (1933).
\item{$^{63}$} W.H. Barkas, Phys.Rev. {\bf 89}, 1019 (1953); W. H.
Barkas, F.M. Smith and W. Birnbaum, Phys. Rev. {\bf 98}, 605 (1955).
\item{$^{64}$} R.M. Sternheimer, Phys.Rev. {\bf 88}, 851 (1952);
Phys.Rev. {\bf 91}, 256 (1953).
\item{$^{65}$} J.S. Ball and G.F. Chew, Phys.Rev. {\bf 109}, 1385 (1958);
J.S. Ball and J.K. Fulco, Phys.Rev. {\bf 113}, 647 (1959).
\item{$^{66}$} M. Ceschia and A. Perlmutter, Nuovo Cimento {\bf 33}, 578
(1964).
\item{$^{67}$} A. Perlmutter, Phys.Lett. {\bf 4}, 336 (1963).
\item{$^{68}$} M. Friedmann, D. Kessler, A. Levy, A. Perlmutter, Nuovo
Cimento {\bf 35}, 355 (1965).
\item{$^{69}$} J. Cox and A. Perlmutter, Nuovo Cimento {\bf 37}, 761 (1965).
\item{$^{70}$} A. Perlmutter, Physics Today, {\bf 49}, {\bf 84}, August 1998.
\item{$^{71}$} H. Yukawa, Proc.Phys. Math. Soc. Japan {\bf 17}, 48
(1935).
\item{$^{72}$} C.D. Anderson and S. Neddermeyer, Phys.Rev. {\bf 51}, 884
(1937); {\bf 54}, 99 (1938).
\item{$^{73}$} P.A.M. Dirac, Proc. Roy. Soc. {\bf A117}, 610 (1928).
\item{$^{74}$} R. Marshak and H. Bethe, Phys.Rev. {\bf 72}, 506 (1947).
\item{$^{75}$} M. Gell Mann, Phy. Lett.{\bf 8}, 214 (1964); G. Zweig,
CERN Report 8419/Th 412, 1964.
\item{$^{76}$} W. Heisenberg, Z. Physik {\bf 77}, 1 (1932).
\end